\documentclass[twocolumn,showpacs,preprintnumbers,amsmath,amssymb]{revtex4}
\usepackage[dvips]{graphicx}
\usepackage{dcolumn}

\begin{document}

\title
{Stability of Half-Quantum Vortices in Equal-Spin Pairing States of $^3$He}

\author{Natsuo Nagamura and Ryusuke Ikeda}

\affiliation{%
Department of Physics, Kyoto University, Kyoto 606-8502, Japan
}

\date{\today}

\begin{abstract} 
Recent experiments on superfluid $^3$He in globally anisotropic aerogels have shown realization of the polar superfluid phase and of the half-quantum vortices (HQVs) in this phase upon rotation. To clarify why the HQVs, which had not been detected clearly in the A phase of the bulk liquid, have been realized in the polar phase, we theoretically examine the relative stability of a HQV-pair against a single phase vortex in both the bulk A-phase and the polar phase in an aerogel. By taking care of important roles of a higher order gradient term, which assists the stability of HQVs but has never been incorporated so far in the Ginzburg-Landau (GL) approach, we find that several consequences, including the extension of the polar phase at {\it lower} pressures in the phase diagram, facilitate realization of the HQVs there in contrast to the case of the bulk A phase in a slab 
geometry. 
\end{abstract}



\maketitle

\section{Introduction}
\label{sec:intro}

　Recent experimental works on superfluid $^3$He in strongly anisotropic (or, one dimensional) aerogels have shown the presence of the polar pairing state \cite{Dmitriev15} arising from an anisotropy-induced lift \cite{AI06} of degeneracy between different pairing states and have found a clear evidence of realization of the long-sought half-quantum vortices (HQVs) in the rotating polar phase \cite{Autti16,Bil18,Jim18}. So far, HQVs have been searched for in the A-phase of superfluid $^3$He in a slab geometry and by expecting the orbital ${\bf l}$-vector to be locked to the surface normal of the film plane. The fixed ${\bf l}$-vector to the surface normal implies that the orbital degrees of freedom of the Cooper pair condensate are frozen out. In spite of intensive studies performed so far, no convincing data suggesting the presence of HQVs have been reported in $^3$He-A phase. 

The seminal theoretical work performed in the London limit has indicated \cite{SV} that a pair of HQVs is always lowered in energy compared to a single phase-vortex (PV) with the uniform ${\bf d}$-vector as far as the ${\bf l}$-vector is locked to the surface normal. In the London limit, the gradient energy of one HQV-pair becomes lower than that of a PV with increasing the distance $a$ between the two HQVs in the pair, and the stable size of the pair is determined by balancing this gain of the gradient energy with the dipole energy increasing with $a$ \cite{SV}. Based on these results in the London limit, the absence of clear evidence of HQVs in $^3$He-A in a slab geometry is often ascribed to an experimental problem such that the film thickness is too large to lock the ${\bf l}$-vector to the surface-normal of the film 
plane. 

On the other hand, numerical works on the {\it conventional} Ginzburg-Landau (GL) equations performed more recently have suggested \cite{Machida,Nakahara,Mizushima} that the strong-coupling (SC) corrections to the condensation energy tend to destabilize a HQV relative to a PV. Since no effect of SC corrections is incorporated in the London limit used in Ref.\cite{SV}, the London limit might have overestimated the stability of HQVs. However, a HQV {\it pair} has not been considered in the conventional GL analysis, and hence, the previous results \cite{Machida,Nakahara} cannot be directly compared with the result \cite{SV} in the London limit. In fact, the pair size dependence stabilizing a HQV pair in the London limit occurs from the Fermi liquid (FL) correction to the gradient energy \cite{Leggett,Cross} which is not incorporated at all in the conventional GL model. 

  Further, the experimental fact that the HQV has been seen in the polar phase in aerogels, while it has not in the bulk A-phase in a slab geometry needs to be clarified. 
In considering HQVs in the A-phase, we focus on the situation in a slab geometry under a weak magnetic field perpendicular to the film plane. We assume that the film thickness is thin enough to make the dipole energy ineffective and that the ${\bf d}$-vector has been confined to be parallel to the film plane by the magnetic field \cite{SV}. Then, the issue on HQVs in the A-phase in a slab geometry can be considered on the same ground as that in the polar phase. A clear difference between the two cases is that the HQV in the latter is a line object. Then, a question arises that the HQV in the polar phase might not be intrinsically stable but be stabilized just by a strong pinning to the one-dimensional (line-like) aerogel. 

In this work, we examine stability of a HQV pair in the bulk A phase and the corresponding issue in the polar phase realized in anisotropic aerogels on the same footing. To perform this, the GL theory for the superfluid $^3$He needs to be extended in a form enabling one to study stability of a HQV pair relative to a single PV by clarifying how the FL correction to the gradient terms in the London limit stemming, together with the so-called SC corrections to the bulk free energy, from the repulsive channel of the interaction between the quasiparticles should appear in microscopically deriving the GL free energy. To examine the gradient energy terms beyond the conventional weak-coupling description, we use two models, the FL model based on the use of the four-point vertex between the normal quasiparticles in the FL theory \cite{AGD} and the spin fluctuation (SF) model describing the quasiparticle effective interactions in terms of the SF propagator in the same manner as the derivation of the so-called SC corrections to the bulk GL free energy \cite{BSA}. We find new terms consistent with the FL-corrected gradient terms in the London limit in both the two approaches mentioned above. Performing numerical analysis of the variational equations of the resulting extended GL free energy, we find that, in the bulk A phase, a HQV pair may be intrinsically stabilized far below the superfluid transition, and that, in the polar phase in anisotropic aerogels, the HQV pair is certainly stabilized even near the polar to normal transition temperature where the SC correction, which is unfavorable for the HQV's stability, becomes the most important. Therefore, it is argued that the emergence of the HQVs in the polar phase \cite{Autti16} is not due to the strong pinning of the vortices to the one-dimensional aerogel structure. Further, we find that these results on the HQV-pair's stability in the superfluid $^3$He phases in the two different situations do not depend much on the detail of the effective interaction between the quasiparticles. 
Our results definitely show that the HQVs in the bulk A phase are less stable than those in the polar phase. More or less, this is partly due to the fact that the bulk A phase occurs only in the high pressure region where the SC correction to the bulk free energy becomes important. The competition in the bulk A phase between the gradient energy and the bulk SC free energy term will lead to emergence of a single PV as a metastable defect and hence, tends to result in a coexistent state of the HQVs and the PVs there if experimentally entering the superfluid phase via a rapid cooling upon rotation. 

This paper is organized as follows. We review a basic model and the familiar weak-coupling results on the bulk GL free energy terms in sec.II and the results on the HQV pair in the bulk A phase in London limit in sec.III. In sec.IV, the SF model on the effective interaction between the quasiparticles is reviewed, and the presence of a nontrivial term is pointed out. In sec.V and VI, the gradient energy terms of the quartic order in the order parameter are carefully examined to find out the terms asisting the stability of a HQV pair. In sec.VII, the vortex core energies of a HQV pair and a PV are evaluated to qualitatively point out what affects the stability of a HQV pair. Our numerical results on the variational equations on the extended GL free energies obtained by incorporating the new terms are presented in sec.VIII. Summary of our results is given in sec.IX. Details of theoretical calculations which are needed to obtain new GL terms are shown in Appendix. 

\section{Review on Pairing States}

The superfluid order parameter is defined from the off-diagonal average $\Delta_{\alpha \beta}({\bf p};{\bf k}) = \langle a_{{\bf p}+{\bf k}/2, \alpha} a_{-{\bf p}+{\bf k}/2, \beta} \rangle$ in the form 
\begin{equation}
\Delta_{\alpha \beta}({\bf p};{\bf k}) = i (\sigma_2 \sigma_\mu)_{\alpha \beta} D_\mu({\bf p};{\bf k}) 
\label{eq:OP1}
\end{equation}
with 
\begin{equation}
D_\mu({\bf p};{\bf k}) = A_{\mu,i}({\bf k}) {\hat p}_i. 
\label{eq:OP2}
\end{equation}
Here, ${\bf k}$ means the center-of-mass momentum of the Cooper-pair. Further, the amplitude $|\Delta|$ of the order parameter is conventionally defined by $|\Delta|^2 = \langle A_{\mu,i}^* A_{\mu,i} \rangle$. 
Then, up to O($|\Delta|^4$), the bulk energy contribution of the GL free energy takes the form  
\begin{eqnarray}
F_{bulk} &=& \int d^3r \biggl[ \alpha(T) A_{\mu j}^* A_{\mu j} + \beta_1 |A_{\mu i}A_{\mu i}|^2 + \beta_2 (A_{\mu i}A_{\mu i}^*)^2 \nonumber \\
&+& \beta_3 A_{\mu i}^*A_{\nu i}^*A_{\mu j}A_{\nu j} + \beta_4 A_{\mu i}^*A_{\nu i} A_{\nu j}^* A_{\mu j} \nonumber \\
&+& \beta_5 A_{\mu i}^*A_{\nu i}A_{\nu j}A_{\mu j}^* \biggr]. 
\label{eq:GLfb}
\end{eqnarray}
In the weak-coupling (WC) approximation, the five quartic terms in eq.(\ref{eq:GLfb}) occur from Fig.1, and the corresponding value of each $\beta_j$, which will be denoted as $\beta_j^{({\rm wc})}$, satisfies the relations 
\begin{eqnarray}
\beta_3^{({\rm wc})} &=& -2 \beta_1^{({\rm wc})}, \\ \nonumber 
\beta_2^{({\rm wc})} &=& \beta_4^{({\rm wc})} = -\beta_5^{({\rm wc})}. 
\label{eq:WCbeta0}
\end{eqnarray}
In the bulk liquid, $\beta_2^{({\rm wc})}=\beta_3^{({\rm wc})} = 2 \beta_0 \equiv N(0) |\psi^{(2)}(1/2)|/(240 \pi^2 T^2)$. 

\begin{figure}[t]
\scalebox{0.6}[0.6]{\includegraphics{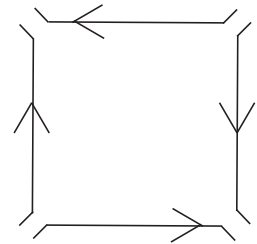}}
\caption{"Gor'kov box" expressing the quartic terms of the GL free energy in the WC approximation.}
\label{fig.1}
\end{figure}

The order parameter in the A-phase is of the form 
\begin{equation}
A_{\mu j} = \frac{\Delta}{\sqrt{2}} {\hat d}_\mu ({\hat m} + {\rm i} 
{\hat n})_j, 
\label{eq:opa1}
\end{equation}
where each vector is a real unit vector, and ${\hat m}$ and ${\hat n}$ are orthogonal to each other. Hereafter, we often consider the A phase with ${\bf l} = {\bf m} \times {\bf n}$ fixed to ${\hat z}$ far from a vortex core. Then, 
\begin{equation}
A_{\mu j} = \frac{|\Delta|}{\sqrt{2}} e^{{\rm i} \Phi} {\hat d}_\mu ({\hat x} + {\rm i} {\hat y})_j 
\label{eq:opa2}
\end{equation}
can be used in place of eq.(\ref{eq:opa1}) far from a vortex core. On the other hand, the order parameter in the polar phase takes the form 
\begin{equation}
A_{\mu,j} = |\Delta| e^{{\rm i} \Phi} {\hat d}_\mu 
{\hat z}_j. 
\label{eq:opp}
\end{equation}
By substituting the order parameters listed above into $F_{bulk}$, the coefficient of the $|\Delta|^4$ term in $F_{bulk}$ is determined depending on the pairing symmetry. For instance, between the corresponding coefficients, $\beta_A=\beta_2+\beta_4+\beta_5$, $\beta_B=\beta_1+\beta_2+(\beta_3+\beta_4+\beta_5)/3$, $\beta_{pol}=\sum_{j=1, \cdot\cdot\cdot, 5} \beta_j$, and $\beta_P=\beta_1+\beta_2+(\beta_3+\beta_4+\beta_5)/2$, for the A, B, polar, and planar phases, respectively, the following relation is found to be satisfied 
\begin{equation}
\beta_A^{({\rm wc})} : \beta_B^{({\rm wc})} : \beta_{pol}^{({\rm wc})} : \beta_P^{({\rm wc})} = 6 : 5 : 9 : 6 . 
\label{eq:betaratio} 
\end{equation} 
According to the conventional mean field theory for an ordered phase near a second order transition, the most stable pairing state of the bulk liquid $^3$He has the lowest value of the coefficient of the quartic term. Thus, according to eq.(\ref{eq:betaratio}), the B phase is always realized in the bulk liuid in equilibrium in the WC approximation. The well-known SC contributions to $\beta_j$ which stabilize the A phase at higher pressures will be mentioned in sec.IV. 

\section{HQV in London limit}
\label{sec:L}

 First, let us start from reviewing the gradient energy in the situation \cite{SV} with one HQV-pair. As far as equal-spin-pairing states with frozen orbital components are concerned, the gradient energy in the London limit is commonly expressed as 
\begin{equation}
F_L/L_z = \frac{1}{2} \int d^2r [ {\cal K}_s (\nabla \Phi)^2 + {\cal K}_{sp} \nabla d_\mu \cdot \nabla d_\mu \biggr]. 
\label{eq:FL1}
\end{equation}
The phase $\Phi$ and the ${\bf d}$-vector $d_\mu$ in the case of one HQV pair are represented by 
\begin{eqnarray}
\Phi &=& \frac{1}{2}(\varphi_++\varphi_-), \nonumber \\
{\bf d} &=& {\hat e}_x \, {\rm cos}\alpha + {\hat e}_y \, {\rm sin}\alpha
\label{eq:phihq}
\end{eqnarray}
with $\varphi_\pm = {\rm tan}^{-1}[(y/(x \mp a/2)]$, and $\alpha = (\varphi_+ - \varphi_-)/2$, where ${\hat e}_x$ is the unit vector in $x$-direction, and $a$ is the size of a HQV-pair (see Fig.2). In this case, eq.(\ref{eq:FL1}) is given by \cite{SV} 
\begin{equation}
F_L(a)/L_z = \pi {\cal K}_s {\rm ln}\biggl(\frac{r_\Omega}{\xi_c} \biggr) - \frac{\pi}{2} ({\cal K}_s - {\cal K}_{sp}){\rm ln}\biggl(\frac{a}{\xi_c} \biggr),
\label{eq:FL2}
\end{equation}
respectively, where a lower cut-off length $\xi_c$ of the order of the coherence length is assumed, and the upper cut off length $r_\Omega$ may be, as usual, identified with the average spacing between neighboring HQV-pairs determined by the rotation angular velocity $\Omega$. The first term of eq.(\ref{eq:FL2}) is nothing but the energy $F_L^{({\rm PV})}$ of a single PV, and the presence of the second term with ${\cal K}_s > {\cal K}_{sp}$ \cite{Cross,SV} indicates the stability of a HQV pair in the London limit where $a \geq \xi_c$ is assumed. This is why the HQV pair has a lower energy than a PV in the London limit. 

The actual size of the HQV pair is determined in the London limit by minimizing the sum of the gradient energy shown above and the dipole energy w.r.t. the pair size $a$. The vortex energy in the GL approach includes the free energy contributions other than the gradient energy which may arise from various components of the order parameter $A_{\mu,i}$. On the other hand, the dipole energy contribution is effective at long length scales of the order of the dipole coherence length $\xi_d$ where the London limit is safely valid. Therefore, it will be sufficient to clarify the sign of the energy difference between the HQV-pair and a single PV, which corresponds to the above-mentioned $F_L(a)-F_L^{({\rm PV})}$, at length scales comparable with $\xi_d$ in order to judge the stability of a HQV pair relative to a single PV. 

\begin{figure}[t]
\scalebox{0.55}[0.55]{\includegraphics{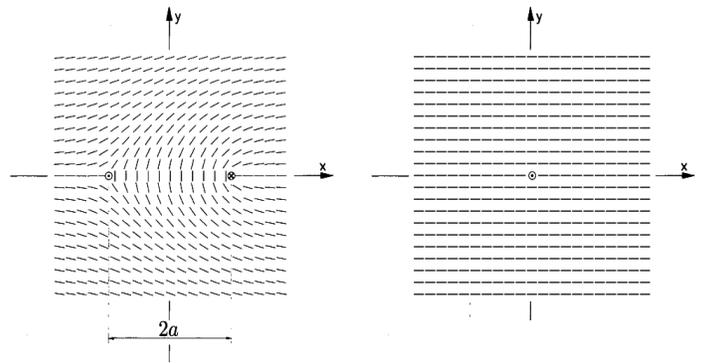}}
\caption{Schematic pictures expressing the textures of $d_\mu$ around one pair of HQVs (left) and a PV (right). }
\label{fig.2}
\end{figure}
 When the conventional FL correction is incorporated, we have the relation \cite{VW} 
\begin{eqnarray}
{\cal K}_s - {\cal K}_{sp} &=& \frac{N(0)}{900}\biggl(\psi^{(2)}(1/2)\frac{|\Delta|^2}{\pi T} \biggr)^2 \nonumber \\ 
&\times& \biggl(\frac{v_{\rm F}}{2 \pi T} \biggr)^2 \biggl[ \frac{F_1^s}{1+F_1^s/3} - \frac{F_1^a}{1+F_1^a/3} \biggr] 
\label{eq:difK1}
\end{eqnarray}
valid up to O($|\Delta|^4$), where $N(0)$ and $v_{\rm F}$ are the DOS and the Fermi velocity of the quasiparticle in the normal state, and $F_{1}^s$ and $F_{1}^a$ are the Landau parameters of the normal Fermi liquid (FL) written in the standard notation and satisfying the relation $F_{1}^s > F_{1}^a$. This expression (\ref{eq:difK1}) has several consequences: First of all, the stability of a HQV pair is guaranteed by the repulsive interaction between the quasiparticles which does not appear in the conventional WC BCS model taking account only of the attractive component of the interaction. Secondly, the stability of a HQV pair is determined, in the language of the GL theory, by a gradient energy contribution in the nonlinear (O($|\Delta|^4$)) term which are neglected in the conventional GL approach \cite{Machida,Nakahara}. Since, originally, the London limit and the GL theory are two different limits of a single theory, a recovery of the London results is expected by performing an extension of the GL approach. Based on these motivations, a derivation of the second term of eq.(\ref{eq:FL2}) in the GL approach will be considered in sec.VI. 

\begin{figure}[t]
\scalebox{0.5}[0.5]{\includegraphics{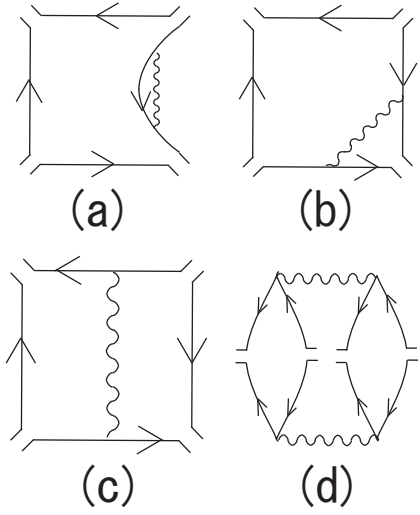}}
\caption{SC diagrams contributing to O($|\Delta|^4$) terms in the GL free 
energy. The wavy line expresses the normal SF propagator $1/(1 - I \chi_N)$. See the text for further details. }
\label{fig.3}
\end{figure}

\section{Spin fluctuation Model of Effective Interaction}

Below, the spin fluctuation approach (SF) will be used as one model to incorporate effects of the repulsive interactions between the quasiparticles. This approach has been used as a simplified model describing the strong coupling (SC) correction to the bulk free energy necessary to stabilize the A-phase at higher pressures and start from defining the bare Hamiltonian for Fermions in the form 
\begin{equation}
{\cal H}_e - \mu N = \sum_{{\bf p},\sigma} \xi_{\bf p} a^\dagger_{{\bf p},\sigma} a_{{\bf p},\sigma} + I \sum_{{\bf p}, {\bf p}'} a^\dagger_{{\bf p},\uparrow} a_{{\bf p},\uparrow} a^\dagger_{{\bf p}',\downarrow} a_{{\bf p}',\downarrow} 
\label{eq:stoner}
\end{equation}
with $I > 0$. 
In this model, the normal SF propagator plays the role of the effective interaction between the quasiparticles. By treating the SF in the Gaussian approximation, the O($|\Delta|^4$) diagrams arising from the SF free energy $F_{fl}$ are described in Fig.3 and expressed by \cite{BSA,RI15,com0} 
\begin{eqnarray}
F_{fl} &=& - \frac{{\overline I}^2}{2 N(0)} T \sum_\Omega \int_{\bf q} \frac{1}{1 - I \chi_N({\bf q}, \Omega)} \delta \chi_{\alpha, \alpha}({\bf q}, \Omega) \nonumber \\
&-& \frac{I^2}{4} T \sum_\Omega \int_{\bf q} \biggl(\frac{1}{1 - I \chi_N({\bf q}, \Omega)} \biggr)^2 \nonumber \\
&\times& \delta \chi_{\alpha, \beta}({\bf q}, \Omega) \delta \chi_{\beta, \alpha}({\bf q}, -\Omega), 
\label{eq:BSA}
\end{eqnarray}
where ${\overline I}=N(0) I$, $\chi_{\alpha,\beta}$ is the spin density correlation function, and $\delta \chi_{\alpha,\beta} = \chi_{\alpha,\beta} - \delta_{\alpha,\beta} \chi_N$ with $\chi_N=\chi_{\alpha, \alpha}|_{\Delta=0}/2$. 

Here, let us examine how the two terms in $F_{fl}$ contribute to $F_{bulk}$, eq.(\ref{eq:GLfb}), in the case of the ordinary bulk liquid. The so-called SC correction to $\beta_j$, $\beta_{j,{\rm se}}^{({\rm sc})}$, which is of O($ \beta_j^{(wc)} T_c/E_{\rm F}$) ($j=1$ to $5$), has followed from the last term of eq.(\ref{eq:BSA}) accompanied by two (normal) SF propagators. One example of the corresponding diagrams is expressed by Fig.3 (d). In the case of the bulk liquid, $\beta_{j,{\rm se}}^{({\rm sc})}$s are given by 
\begin{eqnarray}
\beta^{({\rm sc})}_{1,{\rm se}} &=& -0.1 \beta_0 \delta = \frac{\beta^{({\rm sc})}_{5,{\rm se}}}{7}, \,\,\,\,\,\,\,\, \beta^{({\rm sc})}_{2,{\rm se}} = 0.2 \beta_0 \delta, \nonumber \\
\beta^{({\rm sc})}_{3,{\rm se}} &=& \frac{\beta^{({\rm sc})}_{2,{\rm se}}+5\beta^{({\rm sc})}_{1,{\rm se}}}{6}= \beta^{({\rm sc})}_{4,{\rm se}} - 5 \beta^{({\rm sc})}_{1,{\rm se}}, 
\label{eq:SCbeta0}
\end{eqnarray}
where $\delta \propto T/E_{\rm F}$ was defined in Ref.\cite{BSA}. 
On the other hand, the contribution to the first (quadratic) term of $F_{bulk}$ following from the first term of eq.(\ref{eq:BSA}) can be absorbed into the coefficient $\alpha(T)$ by redefining $T_c$. Further, the contributions of Fig.3 (a) and (b) to the quartic terms of $F_{bulk}$ are found to take the same form as that of the WC diagram Fig.1, and, for this reason, these two diagrams can be regarded as having been absorbed into the $\beta_j$s listed in eq.(4). 

In contrast, much attention should be paid to roles of Fig.3 (c). In Ref.\cite{SR}, the contribution of this diagram to $F_{bulk}$ has been argued to be of a higher order in $T_c/E_{\rm F}$ by neglecting the frequency dependence only in the four-point vertex so that this figure is rather included in the family of Fig.3 (d). Here, this argument will be reconsidered by assuming that the SF in eq.(\ref{eq:BSA}) carries only low energy fluctuations. When following the treatment used in Ref.\cite{RI15}, Fig.3 (c) in the case with a spacially uniform order parameter is expressed in the form 
\begin{eqnarray}
F_{fl}^{(c)} &=& \int_{\bf q} \frac{- 4 \pi T^2 {\overline I}^2}{1 - I \chi_N({\bf q},0)} \sum_\varepsilon \frac{1}{(2|\varepsilon|)^3} \biggl\langle \frac{1}{({\bf v}_{\bf p}\cdot{\bf q})^2 + 4 \varepsilon^2}\nonumber \\
&\times& [2(D_{\mu 0}^*({\bf p}) D_{\mu 0}({\bf p}))^2 + |D_{\mu 0}({\bf p})D_{\mu 0}({\bf p})|^2] \biggr\rangle_{\hat {\bf p}}, 
\label{eq:3c}
\end{eqnarray}
which is nonvanishing and of the same order in $T_c/E_{\rm F}$ as that of Fig.1, where $D_{\mu 0}({\bf p}) = A_{\mu i} {\hat p}_i$, and $\langle \,\,\,\,\rangle_{\hat {\bf p}}$ denotes the angle average over the ${\bf p}$-direction on the Fermi surface. In eq.(\ref{eq:3c}), the quantum ($\Omega \neq 0$) components of the SF was neglected following Ref.\cite{BSA}, and ${\overline I}$ was assumed to be a quantity of the zeroth order in $T_c/E_{\rm F}$. Then, the resulting contribution of Fig.3(c) to each $\beta_j$, $\Delta \beta_j^{(c)}$, satisfies the ratio, $\Delta \beta_1^{(c)} : \Delta \beta_2^{(c)} : \Delta \beta_3^{(c)} : \Delta \beta_4^{(c)} : \Delta \beta_5^{(c)} = 1 : 2 : 2 : 2 : 2$, and, when each $\beta_j^{({\rm wc})}$ is replaced by the sum of $\beta_j^{({\rm wc})} + \Delta \beta_j^{(c)}$, the ratio between the coefficients of the quartic terms for the main $p$-wave pairing states, A, B, polar, and planar pairing states, is found to be the same as the WC one, eq.(\ref{eq:betaratio}). Therefore, the conclusion on the relative stability between the different main pairing states is unaffected by taking account of Fig.3 (c), although the overall value $\beta_0$ should be slightly renormalized by including the contribution of Fig.3 (c). For this reason, we will keep using $\beta_j^{({\rm wc})}$ below as the $\beta_j$ value at the zeroth order both in $T_c/E_{\rm F}$ and the disorder strength. 

\section{Gradient Energy in WC approximation}

Now, the gradient terms will be considered within the GL approach. In the conventional GL approach, the gradient term is taken within the O($|\Delta|^2$) terms  and, in the bulk liquid which is isotropic in real space, has the form 
\begin{equation}
F_{grad2}= \frac{1}{2} \int d^3r [ K_1 \partial_i A_{\mu,j}^* \partial_i A_{\mu,j} + 2 K_2 (\nabla \cdot {\bf A}_{\mu}^*) (\nabla \cdot {\bf A}_\mu) ], 
\label{eq:gradquad}
\end{equation}
where $\nabla \cdot {\bf A}_\mu \equiv \partial_j A_{\mu,j}$. In the case of superfluid $^3$He in an anisotropic aerogel, additional O($|\Delta|^2$) terms occur (see Appendix). However, it is easily verified by fixing $|\Delta|$ and using $d_\mu \delta d_\mu=0$ that the second term of eq.(\ref{eq:FL2}) stabilizing a HQV pair does not occur from any quadratic (i.e., O($|\Delta|^2$) ) gradient terms. This implies that, if using the {\it conventional} GL approach under a fixed $|\Delta|$, the energy of a HQV pair is estimated to be the same as that of a PV \cite{Mineev}. Therefore, to study the stability of a HQV pair against a PV in the GL framework, additional gradient terms have to be searched for in the O($|\Delta|^4$) terms. 

Next, it will be pointed out that the terms like the second term of eq.(\ref{eq:FL2}) stabilizing a HQV pair do not occur within the WC approximation unaccompanied by any repulsive interactions between the quasiparticles. In this WC approximation, the O($|\Delta|^4$) terms occur from the familiar "Gor'kov box" diagram, Fig.1. When the pair-field $\Delta_{\alpha \beta}({\bf p})$ has a center-of-mass momentum which will be denoted as ${\bf k}_j$ below, Fig.1 can be expressed, e.g., in the form 
\begin{eqnarray}
\sum_{{\bf k}_j}  \delta_{{\bf k}_1+{\bf k}_3, {\bf k}_2+{\bf k}_4}\!\!&\langle& \!\! f({\bf v}_{\bf p}\cdot{\bf k}_j)  {\rm Tr}[\Delta({\bf p}; {\bf k}_1) \Delta^\dagger({\bf p}; {\bf k}_2) \nonumber \\
&\times& \Delta({\bf p}; {\bf k}_3) \Delta^\dagger({\bf p}; {\bf k}_4)] \, \, \rangle. 
\label{eq:gradwc1}
\end{eqnarray}
The original expression of the function $f$ will be given in Appendix. The term in which all ${\bf k}_j$ in $f$ are zero results in the quartic terms of $F_{bulk}$. By focusing on the terms of the quadratic order in ${\bf k}_j$ and changing from the momentum (${\bf k}_j$) representation to the center-of-mass coordinate (${\bf r}$) representation, eq.(\ref{eq:gradwc1}) is found to be expressed as a linear combination of the following two kinds of terms,  
\begin{eqnarray}
A_1 \!\! &=& \!\! \int_{\bf r} {\rm Tr}(\sigma_\mu \sigma_\nu \sigma_\rho \sigma_\lambda) \langle p_a p_b (D_\mu^*({\bf p})  \partial_a D_\nu({\bf p}) \partial_b D_\rho^*({\bf p}) \nonumber \\ 
&\times& D_\lambda({\bf p}) 
+  \partial_a D_\mu^*({\bf p}) \partial_b D_\nu({\bf p}) D_\rho^*({\bf p}) D_\lambda({\bf p})) \rangle, \nonumber \\
A_2 \!\! &=& \!\! \int_{\bf r} {\rm Tr}(\sigma_\mu \sigma_\nu \sigma_\rho \sigma_\lambda) \langle p_a p_b (D_\mu^*({\bf p})  \partial_a D_\nu({\bf p}) D_\rho^*({\bf p}) \nonumber \\
&\times&  \partial_b D_\lambda({\bf p}) + \partial_a D_\mu^*({\bf p}) D_\nu({\bf p}) \partial_b D_\rho^*({\bf p}) D_\lambda({\bf p})) \rangle, 
\label{eq:gradwc2}
\end{eqnarray}
where $D_\mu({\bf p})= A_{\mu,i}({\bf r}) {\hat p}_i$. Here, the two pairing states of our interest will be expressed altogether as $e^{{\rm i}\Phi} d_\mu f({\bf p})$ (see eqs.(\ref{eq:opa2}) and (\ref{eq:opp})). Then, $\partial_a D_\mu({\bf p})$ can be written as $e^{{\rm i}\Phi}({\rm i}\partial_a \Phi d_\mu + \partial_a d_\mu) f({\bf p})$. \cite{com1}. By performing the trace over the spin indices and using the relations $d_\mu \nabla d_\mu =0$ and ${\rm Tr}(\sigma_\mu \sigma_\nu \sigma_\rho \sigma_\lambda)=2(\delta_{\mu,\nu} \delta_{\rho,\lambda}+ \delta_{\mu,\lambda} \delta_{\rho,\nu} - \delta_{\mu,\rho} \delta_{\lambda,\nu})$, one can easily verify that, by assuming the amplitude $|\Delta|$ to be fixed, both expressions in eq.(\ref{eq:gradwc2}) are proportional to $(\nabla \Phi)^2 + \nabla d_\mu \cdot \nabla d_\mu$. Thus, nonvanishing contributions to ${\cal K}_s - {\cal K}_{sp}$ (see eq.(\ref{eq:difK1})) do not occur from the expressions in eq.(\ref{eq:gradwc1}). That is, any gradient term stabilizing a HQV pair does not occur at all from Fig.1. 

Clearly, Fig.1 also includes the diagrams with a simple self energy correction like Fig.3 (a). Further, by using the identity ${\rm Tr}(\sigma_\mu \sigma_\nu \sigma_\rho \sigma_\lambda) = {\rm Tr}(\sigma_\mu \sigma_2 \sigma^{\rm T}_\alpha \sigma_2 \sigma_\nu \sigma_\rho \sigma_\lambda \sigma_\alpha)$, the diagrams of the type of Fig.3(b) also take the form of eq.(\ref{eq:gradwc1}). Therefore, we do not have to consider the gradient terms arising from the diagrams of the type of Figs 3 (a) and (b). Then, in the spin-fluctuation approach, the diagram Fig.3 (c) becomes the only diagram contributing to the stability of HQVs up to O($|\Delta|^4$). 

\section{Interaction-induced Gradient Energy in Quartic Order} 

In this section, we will show that the gradient term assisting the stability of the HQV pair corresponding to the second term of eq.(\ref{eq:FL2}) in the London limit occurs from the diagram of the type of Fig.3 (c). As a model of the effective repulsive interaction between the quasiparticles, let us consider two models. One is the conventional Fermi liquid (FL) model, and the other is the SF model reviewed in sec.III. 

\begin{figure}[t]
\scalebox{0.55}[0.55]{\includegraphics{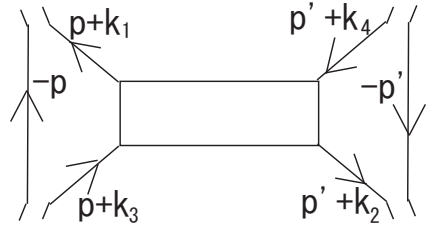}}
\caption{(Color online) Diagram corresponding to Fig.3 (c) rewritten so as to fit to the Fermi liquid description. The rectangle denotes the four point vertex part with a vanishingly small momentum transfer ${\bf k}_1-{\bf k}_3={\bf k}_4-{\bf k}_2$ along the horizontal direction. 
} 
\label{fig.4}
\end{figure}

\subsection{Fermi liquid model}

If we are based on the FL description, the four-point vertex part $\Gamma$, which is the rectangle in Fig.4 corresponding to Fig.3(c), will be assumed to be frequency independent according to Ref.\cite{AGD}. Since the momenta ${\bf k}_j$ carried by the four order parameter fields in Fig.4 are small, $\Gamma$ is assumed to take the form \cite{AGD} 
\begin{eqnarray}
2 N(0)\! &\times& \! \Gamma_{\alpha \beta, \gamma \delta}({\bf p}+{\bf k}_1,{\bf p}'+{\bf k}_2; {\bf p}+{\bf k}_3,{\bf p}'+{\bf k}_4)  \nonumber \\ 
&\simeq&  \Gamma^{(s)}({\hat {\bf p}}\cdot{\hat {\bf p}'}) \delta_{\alpha, \gamma} \delta_{\beta,\delta} + \Gamma^{(a)}({\hat {\bf p}}\cdot{\hat {\bf p}'}) (\sigma_s)_{\alpha, \gamma} (\sigma_s)_{\beta,\delta} \nonumber \\
\label{eq:VC}
\end{eqnarray}
under the momentum conservation ${\bf k}_1+{\bf k}_2={\bf k}_3+{\bf k}_4$, where 
\begin{equation}
\Gamma^{(u)}({\rm cos}\theta) = \sum_{l \geq 0} \Gamma_l^{(u)} P_l({\rm cos}\theta)
\label{eq:VCpw}
\end{equation}
($u=s$, $a$), $\Gamma_l^{(u)}=F_l^{(u)}/(1+F_l^{(u)}/(2l+1))$ with the Landau parameters $F_l^{(u)}$, and $P_l(x)$ is the Legendre polynomial. 
The resulting free energy term is expressed in the form
\begin{widetext}
\begin{eqnarray}
F_{FL4} 
&=& \sum_{k_j} \delta_{{\bf k}_1+{\bf k}_2, {\bf k}_3+{\bf k}_4} \int_{\bf p} \int_{{\bf p}'} \Gamma_{\alpha \beta, \gamma \delta}({\bf p}+{\bf k}_1,{\bf p}'+{\bf k}_2; {\bf p}+{\bf k}_3,{\bf p}'+{\bf k}_4)  T \sum_\varepsilon {\cal G}_{p+k_1}(\varepsilon) {\cal G}_{-p}(-\varepsilon) {\cal G}_{p+k_3}(\varepsilon) \nonumber \\
&\times& T \sum_{\varepsilon'}{\cal G}_{p'+k_2}(\varepsilon') {\cal G}_{-p'}(-\varepsilon') {\cal G}_{p'+k_4}(\varepsilon')  (\Delta^\dagger({\bf p}, {\bf k}_1) \Delta({\bf p}, {\bf k}_3))_{\alpha \gamma} (\Delta^\dagger({\bf p}', {\bf k}_2) \Delta({\bf p}', {\bf k}_4))_{\beta \delta},
\label{eq:FLgrad1}
\end{eqnarray}
\end{widetext}
where ${\cal G}_{\bf p}(\varepsilon)=({\rm i}\varepsilon - \xi_p)^{-1}$ is the Matsubara Green's function. Noting that the $\xi_p$-integral of ${\cal G}_{p+k_1}(\varepsilon) {\cal G}_{-p}(-\varepsilon) {\cal G}_{p+k_3}(\varepsilon)$ is $-2\pi{\bf v}\cdot({\bf k}_1+{\bf k}_3)/(2|\varepsilon|)^3$ when the particle-hole symmetry is assumed, this diagram is found to contribute not to the bulk quartic term but only to the gradient term of O($|\Delta|^4$). Using the relation 
\begin{widetext}
\begin{equation}
i \int_{{\bf k_1},{\bf k}_3} e^{{\rm i}({\bf k}_1-{\bf k}_3)\cdot{\bf r}} (k_1 
+ k_3)_j \langle {\hat p}_l {\hat p}_j \Delta^\dagger({\bf p};{\bf k}_1)\Delta({\bf p};{\bf k}_3) \rangle_{\hat {\bf p}} = \frac{1}{15} \sigma_\mu \sigma_\nu (f_{\mu \nu, l} - f^*_{\nu \mu, l}),
\end{equation}
where 
\begin{equation}
f_{\mu \nu, l} = A^*_{\mu s} \partial_l A_{\nu s} + A^*_{\mu l} \partial_j A_{\nu j} + A^*_{\mu j} \partial_j A_{\nu l}, 
\label{eq:fmunu}
\end{equation}
and keeping only the $l \leq 1$ components in eq.(\ref{eq:VCpw}), 
we obtain 
\begin{equation}
F_{FLgrad4} = N(0) \biggl(\frac{\psi^{(2)}(1/2)}{120 \pi T} \biggr)^2 \biggl(\frac{v_{\rm F}}{2 \pi T} \biggr)^2 \int_{\bf r} [ 2 \Gamma_1^{(s)} {\rm Im}f_{\mu \mu,l}\cdot {\rm Im}f_{\lambda \lambda,l} + \Gamma_1^{(a)} {\rm Re}(f_{\mu \nu, l} - f_{\nu \mu, l})\cdot{\rm Re}(f_{\mu \nu, l} - f_{\nu \mu, l}) ].
\label{eq:FLgrad2}
\end{equation}

In the London limit, eq.(\ref{eq:FLgrad2}) is highly simplified and, in the A phase with the ${\bf l}$-vector along the rotation axis ($\parallel {\hat z}$), takes the form 
\begin{equation}
F_{FLL4} = \frac{N(0)}{1800} \biggl(\frac{\psi^{(2)}(1/2) |\Delta|^2}{\pi T} \biggr)^2 \biggl(\frac{v_{\rm F}}{2 \pi T} \biggr)^2 \int_{\bf r} [ \Gamma_1^{(s)} (\nabla_\perp \Phi)^2 + \Gamma_1^{(a)} \nabla_\perp d_\mu \cdot \nabla_\perp d_\mu],
\label{eq:FLgradL}
\end{equation}
\end{widetext}
which is nothing but eq.(\ref{eq:FL1}) with eq.(\ref{eq:difK1}) satisfied. In the polar phase, the corresponding $F_{FLL4}$ is the quarter of eq.(\ref{eq:FLgradL}). 

The expression following by substituting eq.(\ref{eq:fmunu}) into eq.(\ref{eq:FLgrad2}) is highly involved. Since the ratio of the Landau parameters $|F_1^{(a)}/F_1^{(s)}|$ is quite small, however, eq.(\ref{eq:FLgrad2}) can be simplified by neglecting the antisymmetric term proportional to $\Gamma_1^{(a)}$. The resulting expression will be given in eq.(\ref{eq:FLgrad4}) and will be used for our numerical analysis. 

\subsection{Spin fluctuation model}

To explain derivation of the result on Fig.3 (c) in this model, it will be sufficient to consider the expressions corresponding to eq.(\ref{eq:gradwc2}) based on eq.(\ref{eq:BSA}) given in sec.III. Noting that $\delta \chi_{\alpha \alpha}({\bf q})$ is accompanied by two Pauli matrices, $\sigma_\alpha$, at the external vertices, the corresponding expressions to eqs.(\ref{eq:gradwc2}) we should examine are 
\begin{eqnarray}
C_1 \!\!\! &=& \!\!\! \int_{\bf r} {\rm Tr}(\sigma_\mu \sigma_\nu \sigma_\alpha \sigma_\rho \sigma_\lambda \sigma_\alpha) \langle p_a p_b (D_\mu^*({\bf p})  \partial_a D_\nu({\bf p}) \partial_b D_\rho^*({\bf p}) \nonumber \\ 
\!\! &\times& \!\!  D_\lambda({\bf p}) 
+ \partial_a D_\mu^*({\bf p}) \partial_b D_\nu({\bf p}) D_\rho^*({\bf p}) D_\lambda({\bf p})) \rangle, \nonumber \\
C_2 \!\!\! &=& \!\!\! \int_{\bf r} {\rm Tr}(\sigma_\mu \sigma_\nu \sigma_\alpha \sigma_\rho \sigma_\lambda \sigma_\alpha) \langle p_a p_b (D_\mu^*({\bf p})  \partial_a D_\nu({\bf p}) D_\rho^*({\bf p}) \nonumber \\ 
\!\!\! &\times& \!\!\! \partial_b D_\lambda({\bf p}) + \partial_a D_\mu^*({\bf p}) D_\nu({\bf p}) \partial_b D_\rho^*({\bf p}) D_\lambda({\bf p})) \rangle. 
\label{eq:gradsc2}
\end{eqnarray}
Using ${\rm Tr}(\sigma_\mu \sigma_\nu \sigma_\alpha \sigma_\rho \sigma_\lambda \sigma_\alpha)=2(3 \delta_{\mu,\nu} \delta_{\rho,\lambda}- \delta_{\mu,\lambda} \delta_{\rho,\nu} + \delta_{\mu,\rho} \delta_{\lambda,\nu})$, one finds that the two expressions of eq.(\ref{eq:gradsc2}) are proportional to $3(\nabla \Phi)^2 + \nabla d_\mu \cdot \nabla d_\mu$ and $-3(\nabla \Phi)^2 + \nabla d_\mu \cdot \nabla d_\mu$, respectively. Thus, a difference between the coefficients of $(\nabla \Phi)^2$ and $\nabla d_\mu \cdot \nabla d_\mu$ terms may occur from Fig.3 (c) depending on the pairing states. The detailed form, composed of fifty-one kinds of invariants, of the gradient term of O($|\Delta|^4$) resulting from Fig.3 (c) is presented in eq.(\ref{eq:Sgrad4}) in Appendix. 

When the London limit is taken in the present SF model, the difference in the coefficient, ${\cal K}_s - {\cal K}_{sp}$, corresponding to eq.(\ref{eq:difK1}) becomes 
\begin{widetext}
\begin{equation}
{\cal K}_s - {\cal K}_{sp} = \frac{N(0)}{2240} \biggl(\frac{|\Delta|^2}{\pi T} \biggr)^2 \biggl(\frac{v_{\rm F}}{2 \pi T} \biggr)^2 [u_{1+} + u_{1-} - 2(u_{2+}+u_{2-}) ]
\label{eq:difK2}
\end{equation}
\end{widetext}
for the state (\ref{eq:opa2}) of the A-phase, where $u_{1\pm}$ and $u_{2\pm}$ are defined in eq.(\ref{eq:upmn}) in Appendix. 

In the polar phase created by an anisotropic aerogel, the corresponding expression to eq.(\ref{eq:difK2}) becomes its half value. In numerically investigating the vortices  not only eqs.(\ref{eq:gradquad}) and (\ref{eq:Sgrad4}) or (\ref{eq:FLgrad4}) but also the anisotropy-induced O($|\Delta|^2$) gradient terms (\ref{eq:WCgradani}) and (\ref{eq:SCgradani}), given in Appendix, will also be incorporated. 

Before ending this section, the numerical factors in eqs.(\ref{eq:difK1}) and (\ref{eq:difK2}) will be compared with each other. If the experimental values of the Landau parameters at 30 bar \cite{VW} are used, eq.(\ref{eq:difK1}) becomes 
\begin{equation}
0.7 N(0) \biggl(\frac{v_{\rm F}|\Delta|^2}{2 \pi^2 T^2} \biggr)^2.  
\label{eq:FLsp}
\end{equation}
On the other hand, eq.(\ref{eq:difK2}) becomes 
\begin{equation}
1.8 N(0) \biggl(\frac{v_{\rm F}|\Delta|^2}{2 \pi^2 T^2} \biggr)^2
\label{eq:ksksp}
\end{equation}
for the bulk liquid when the values ${\overline I} = 0.724$ and $T/E_{\rm F}=1.2 \times 10^{-3}$ are used (See the caption of Fig.5 below). 

Therefore, it is anticipated that the SF model of the repulsive interaction between the quasiparticles stabilizes the HQV pair more easily compared with the FL model. Since the Landau parameters always satisfy $F_1^s \gg |F_1^a|$, and $\Gamma_1^s$ is not so sensitive to the pressure in the pressure range between 5 (bar) and 30 (bar) \cite{VW} as well as ${\overline I}$ \cite{BSA} which is nothing but the parameter that detemines the coefficient $1.8$ in eq.(\ref{eq:ksksp}), the difference in the ${\cal K}_s - {\cal K}_{sp}$ value between the two models may be regarded as being essentially independent of the pressure value. 

\section{Estimation of Vortex Core Energy} 

In the preceding section, the energy gain in a HQV-pair through the gradient energy, which had been suggested in the study in the London limit, has been found by microscopically reexamining the GL approach. On the other hand, the cost of the condensation energy near the vortex core, i.e., the vortex core energy, is also nonnegligible to understand how the HQV is stabilized depending on the pairing states. The dependences on the pairing states arise from the difference in the combination of the $\beta_j$ parameters. To express it clearly in the presence of HQVs in an equal-spin-pairing state, it is convenient to rewrite the quartic terms of the GL free energy in the chirarity basis ${\hat e}_\pm = \mp ({\hat e}_x \pm {\rm i} {\hat e}_y)/\sqrt{2}$ and ${\hat e}_0={\hat e}_z$. Then, the pair field is represented by $A_{ab}$ ($a$, $b = \pm$, $0$) in place of $A_{\mu i}$ ($\mu$, $i = x$, $y$, and $z$). 

First, the case of the bulk A phase with ${\hat {\bf l}} \parallel {\hat z}$ will be considered. The corresponding GL quartic terms are simply expressed by 
\begin{equation}
f_{bulk}^{(4)} = (\beta_2 + \beta_4)|(|A_{++}|^2 + |A_{-+}|^2)^2 + 4 \beta_5 |A_{++} A_{-+}|^2. 
\label{eq:fbulk4}
\end{equation}
The order parameter in the presence of one HQV-pair is expressed in terms of eqs.(\ref{eq:phihq}) by 
\begin{equation}
\frac{A_{\mu,i}}{|\Delta|} = e^{{\rm i} \Phi} d_\mu ({\hat e}_+)_i = \frac{1}{\sqrt{2}} (e^{{\rm i}\varphi_+} ({\hat e}_-)_\mu + e^{{\rm i}\varphi_-} ({\hat e}_+)_\mu) ({\hat e}_+)_i
\label{eq:opa3}
\end{equation}
so that $A_{-+}=|\Delta| e^{{\rm i}\varphi_+}$, and $A_{++} = |\Delta| e^{{\rm i}\varphi_-}$. That is, as is seen later, a pair of HQVs' solution of $A_{\mu,i}$ correspond to those of integer vortices of $A_{\pm +}$, \cite{Machida,Nakahara} (see Figs.2 and 6). Then, the core energy of a pair of HQVs is estimated as $2 (\beta_2+\beta_4) \xi^2$ if setting $|A_{\pm +}|$ far from the vortex cores to be unity and assuming the core area to be $\xi^2$. On the other hand, since both of $A_{\pm +}$ vanish at the core in the case of a single PV, the corresponding one of a single PV is $4 \beta_A \xi^2$. Thus, the HQV obtains a gain in the condensation energy if the ratio 
\begin{equation}
r_A=\frac{\beta_2+\beta_4}{2(\beta_2+\beta_4+\beta_5)}
\label{eq:ratio1}
\end{equation}
is less than unity. In the WC approximation, this ratio is just unity. If, however, the SC correction obeying the relation (\ref{eq:SCbeta0}) is incorporated, this ratio becomes $r_A=(1 + 0.4375 \delta)$. That is, as noted in Ref.\cite{Mizushima}, the HQVs are destabilized by the SC corrections to $\beta_j$. 

Now, we turn to the corresponding quantities in the polar phase in an anisotropic aerogel. In this case, the order parameter in the presence of a HQV pair is expressed by eq.(\ref{eq:opa3}) with $({\hat e}_+)_i$ replaced by $({\hat e}_0)_i$, and the corresponding expression of eq.(\ref{eq:fbulk4}) is given by replacing $A_{\pm +}$, $\beta_2 + \beta_4$, and $\beta_5$ there by $A_{\pm 0}$, $\beta_2+\beta_3+\beta_4$, and $\beta_1+\beta_5$, respectively. In this case, however, there are additional quartic terms induced by the anisotropy parameter $\delta_u$ denoted as $\beta_{jz}^{({\rm wc})}$ in Ref.\cite{AI06} (see Appendix). Then, the resulting ratio $r_{pol}$ corresponding to eq.(\ref{eq:ratio1}) is given by 
\begin{equation}
r_{pol} = \frac{\beta_2+\beta_3+\beta_4+2(\beta_{2z}+\beta_{3z}+\beta_{4z})}{2[\sum_{j=1 \cdot\cdot\cdot 5} (\beta_j + 2 \beta_{jz})]}.
\label{eq:ratio2}
\end{equation}
When the expressions shown in Ref.\cite{AI06} of $\beta_{jz}$ are used, the ratio $r_{pol}$ becomes precisely unity within the WC approximation, and, by incorporating the SC correction to $\beta_j$s, we have 
\begin{equation}
r_{pol}=1 + \frac{\delta}{3(1+c_r)}, 
\label{eq:ratio4}
\end{equation}
where the factor $c_r$ is given in Appendix and is positive for a moderately large stretched anisotropy $\delta_u < -0.1$. That is, in the WC approximation, a HQV pair is estimated to have the same core energy as a single PV, while an inclusion of the SC corrections to the coefficients of the bulk quartic terms makes a HQV pair less stable than a PV, although the energy cost of a HQV pair due to the SC corrections in the polar phase is smaller than the corresponding one in the bulk A phase case. 

Further, in the polar phase, another origin of the gain in the core energy of the HQVs is present. As pointed out in Ref.\cite{RI15}, the self energy diagrams reconstructed by the impurity scatterings leads to a large enhancement of the $\beta_j$ parameters which becomes the origin of the absence of the equilibrium A-phase in the globally isotropic aerogel \cite{Halperin}. Due to the presence of the prefactor $2$ in the denominator of the ratio (\ref{eq:ratio2}), these scattering-induced positive contributions, given in eq.(\ref{eq:SCbetavc}) in Appendix, to $\beta_j$s also result in a reduction of the ratio $r_{pol}$. 

\begin{figure}[t]
\scalebox{0.6}[0.7]{\includegraphics{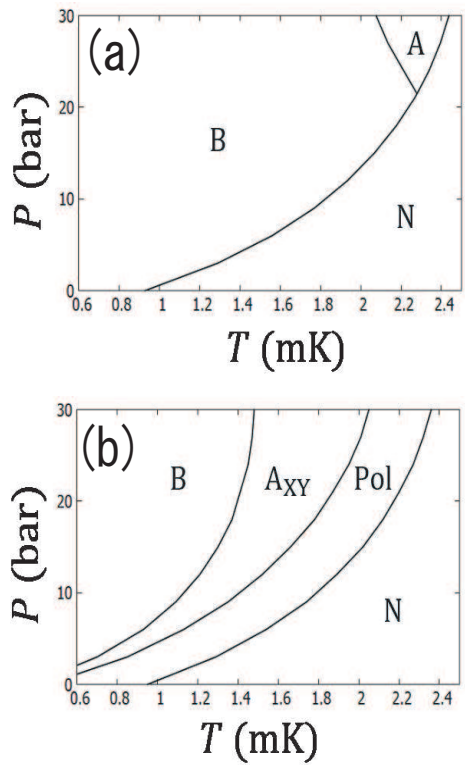}}
\caption{(Color online) Examples of theoretical superfluid phase diagrams of (a) the bulk liquid $^3$He and (b) the liquid $^3$He in a medium with anisotropic and elastic scatting processes which mimics \cite{AI06} an aerogel with a stretched global anisotropy. The figure (a) was obtained in terms of the material parameters given in Ref.\cite{Greywall} and by setting ${\overline I}=0.725$, while the parameter values $(2 \pi \tau)^{-1} = 0.13$ (mK) and the anisotropy parameter defined in Ref.\cite{AI06} $\delta_u=-0.5$ were also used in obtaining the figure (b). The A$_{\rm XY}$ phase denotes the A phase with the ${\bf l}$-vector lying in the plane perpendicular to the anisotropy axis. 
} 
\label{fig.5}
\end{figure}

We have also estimated the ratios $r_A$ and $r_{pol}$ by using the $\beta_j$ parameters determined experimentally for the bulk liquid and the liquid in a globally isotropic aerogel \cite{Choi}. For instance, one finds that using the values in Ref.\cite{Choi} results in $r_A=1.3$ for the bulk $^3$He at 26 bar, which is, according to the expression of $r_A$ given below eq.(\ref{eq:ratio1}), the value corresponding to $\delta=0.7$. This is a reasonable result, judging from the fact that, according to eqs.(4) and (\ref{eq:SCbeta0}), the bulk A phase is stable when $\delta > 0.465$. Similarly, using the corresponding values \cite{Choi} for an isotropic aerogel, one finds that $r_{pol} = 1.01$ and $1.04$ at 5 and 10 bars, respectively. Further, as noted above, a moderately large stretched anisotropy $|\delta_u|$ appearing in $\beta_j$ in the WC approximation through the parameter $c_r$ (see Appendix) seems to further reduce $r_{pol}$. 

One might wonder if, in the case of a highly anisotropic aerogel, anisotropic pairing states are stabilized so that this mechanism explaining the feature in the isotropic aerogel is not reflected there. As shown in Ref.\cite{RI15}, however, the anisotropic pairing states are primarily stabilized by anisotropic SC effects on the {\it quadratic} terms of the GL expansion of the free energy, and the scattering-induced enhancement of $\beta_j$-parameters in the quartic GL terms remains in the case of anisotropic aerogels. In fact, since no additional contributions induced by the anisotropy are incorporated for simplicity in the O($|\Delta|^4$) gradient terms, anisotropic SC contributions to $\beta_j$ \cite{RI15} will not be considered consistently. Therefore, it is concluded that it is the scattering-induced enhancement of $\beta_j$ which becomes one possible origin for stabilizing the HQV in the polar phase. 

\begin{figure}[t]
\scalebox{0.35}[0.35]{\includegraphics{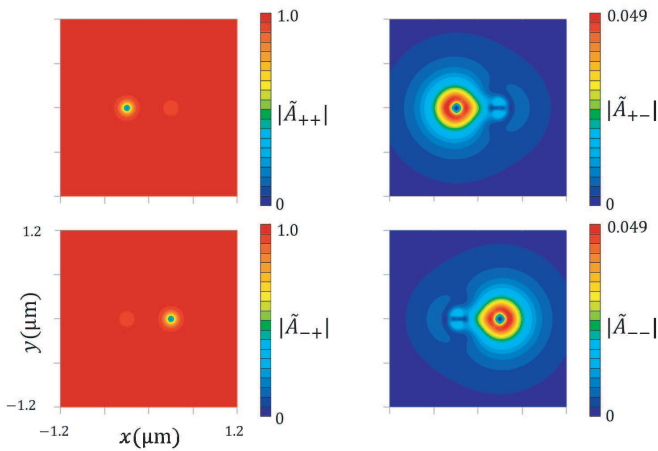}}
\caption{(Color online) Typical example of spatial profiles of the order parameter components $A_{ab}$ (see sec.VII) described on the $x$-$y$ plane in the case of a HQV pair in the bulk A phase. This figure has been obtained consistently with derivation of Fig.7(b).
} 
\label{fig.6}
\end{figure}

\section{Numerical results}

To address whether the HQV is stabilized in the bulk A phase and the polar phase in anisotropic aerogels by using quantitatively reasonable parameters, we have numerically compared the free energy of a HQV pair with that of a single PV in both the systems. 

Both for the bulk A phase and the polar phase in an aerogel, the free energy of a HQV pair is computed by numerically solving variational GL equations following from the extended GL free energy (see below) including the O($|\Delta|^4$) gradient terms, eq.(\ref{eq:Sgrad4}) or (\ref{eq:FLgrad4}), derived in this work. The vortices are assumed to be straight along $z$-axis so that we can focus on the order parameter $A_{\mu,i}(x,y)$ in the $x$-$y$ plane. Further, the longer cutoff length in the $x$-$y$ plane for a single HQV pair is assumed to correspond to the lattice constant of the vortex lattice and to be determined by the magnitude of the rotation velocity \cite{SV}. As a numerical method, we closely follow the direct 2D method in Ref.\cite{Th86} by adopting the London result, eq. (\ref{eq:phihq}), as the outer boundary condition for the order parameter $A_{\mu,i}(x,y)$. Further, the system size in $x$-direction along which the HQV pair with the pair separation $a$ can be extended will be chosen to be ten times longer than that in the perpendicular ($y$) direction. The HQV pair with $a=0$ is nothing but a PV. Therefore, to see the stability of a HQV pair, we examine the free energy of a HQV pair measured from that of a PV, i.e., the free energy difference in equilibrium, $F_{eq}(a) - F_{eq}(a=0)$, as a function of the pair size $a$. 

Strictly speaking, it is necessary, like the analysis in the London limit \cite{SV}, to see the sum of the gradient energy and the dipole energy in order to judge the stability of a HQV pair. However, the dipole energy becomes important only at the large enough scale of the order of 10($\mu$ m) comparable with the dipole coherence length $\xi_d$, and, at such large scales, the results on the dipole energy in the London limit is reliable quantitatively. Therefore, as mentioned in sec.III, we focus here on the $a$-dependence of the vortex energy difference $F_{eq}(a) - F_{eq}(0)$.

\begin{figure}[t]
\scalebox{0.6}[0.6]{\includegraphics{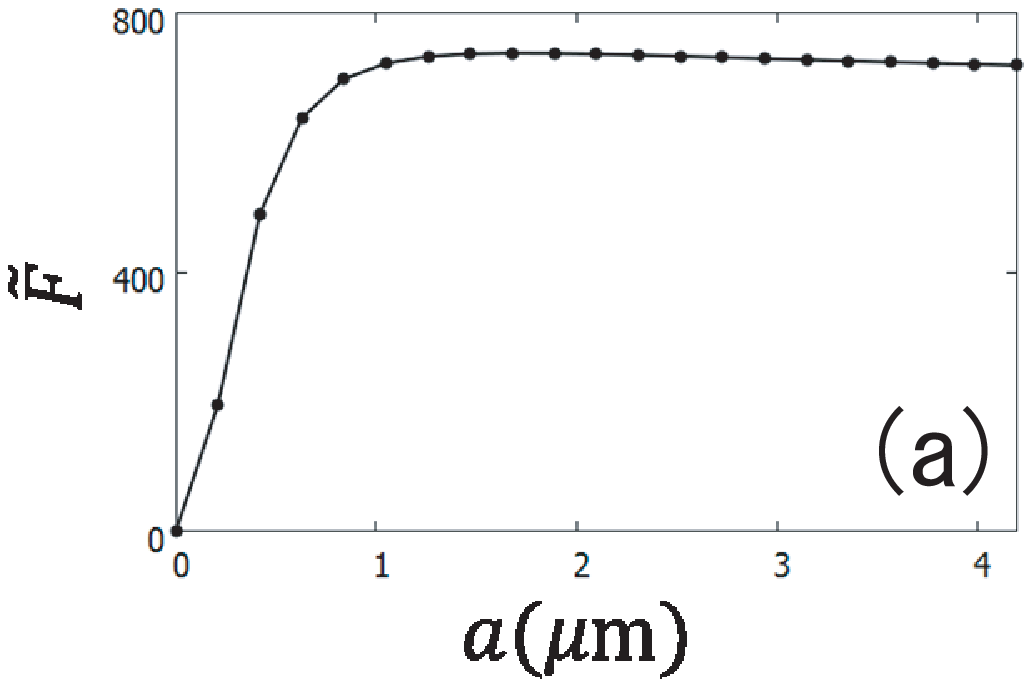}}
\scalebox{0.6}[0.6]{\includegraphics{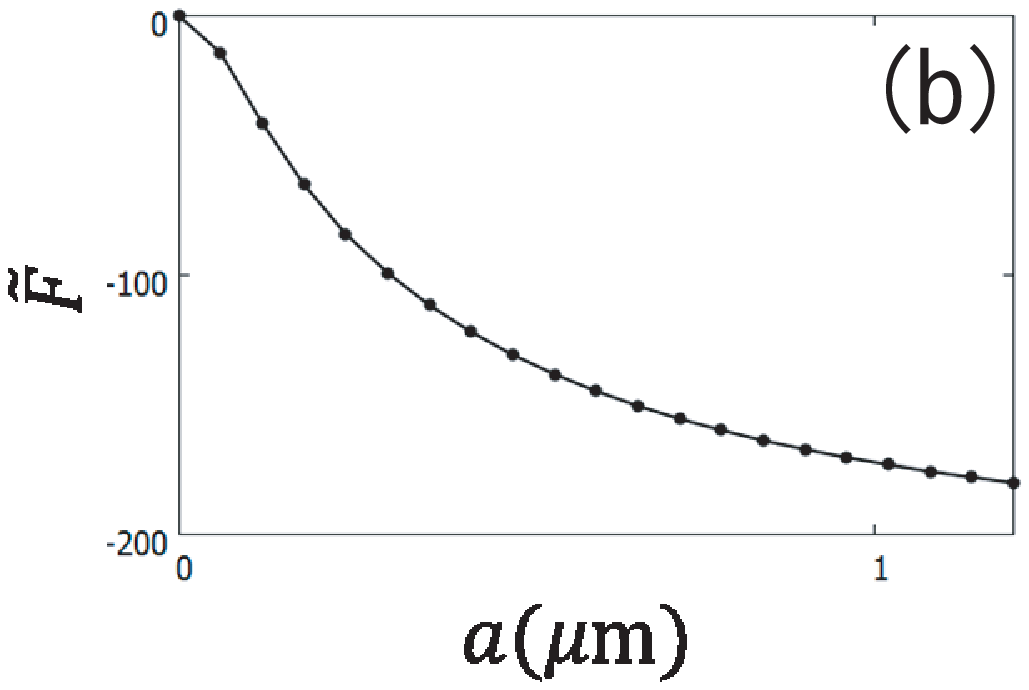}}
\caption{(Color online) Free energy $F_{eq}(a)$ of a HQV pair measured from that of a PV, $F_{eq}(0)$, in the case of the A phase in Fig.5 (a) at 30 bar. The normalization ${\tilde F} \equiv \beta_A(F_{eq}(a) - F_{eq}(0))/\alpha^2(T)$ is used. The figure (a) is ${\tilde F}(a)$ at 2.43 (mK) in the close vicinity of the superfluid transition temperature 2.44 (mK), while the figure (b) is the result at 2.08 (mK) which is just above the A-B transition at 30 bar. The ${\tilde F}$ curve in (a) does not become negative over the lengths of several 
ten ($\mu$m). 
} 
\label{fig.7}
\end{figure}
To determine reasonable parameter values to be used for the numerical computation of the vortices in each phase, we have first started from determining an appropriate pressure ($P$) to temperature ($T$) phase diagram both for the bulk liquid $^3$He and the liquid $^3$He in an aerogel. 
Applying the experimental pressure dependences of $E_{\rm F}$, the bulk $T_c$, and the effective mass $m^*$ of a normal quasiparticle \cite{Greywall} to eqs.(\ref{eq:GLfb}), (4), and (\ref{eq:SCbeta0}), Fig.5 (a) is obtained as the phase diagram of the bulk liquid $^3$He when the interaction parameter ${\overline I}$ is $0.724$, where the relation between ${\overline I}$ and the SC parameter $\delta$ (see sec.IV) given in Ref.\cite{BSA} was used. 

Since, strictly speaking, we focus on the case with a magnetic field parallel to the $z$-axis applied to confine the $d_\mu$-vector to the $x$-$y$ plane, the region of the A-phase can become slightly broader than in the figure. However, we assume that a magnetic field with a moderate magnitude will be sufficient for the in-plane confinement of $d_\mu$ and will not affect the temperature width of the A-phase region.  

On the other hand, in the case of liquid $^3$He in a stretched aerogel, we obtain Fig.5 (b) by applying the values of the scattering strength $1/(2 \pi \tau)=0.13$ (mK) and the dimensionless anisotropy parameter $\delta_u=-0.5$ \cite{AI06} to eqs.(\ref{eq:GLfb}), (4), (\ref{eq:SCbeta0}), (\ref{eq:SCbetavc}), and (\ref{eq:GLfbani}). 

To explain and discuss our results on the vortex energy, we will use our results obtained by using eq.(\ref{eq:FLgrad4}) of the FL model as the O($|\Delta|^4$) gradient energy in most part of this section. Some of the corresponding results following from the use of eq.(\ref{eq:Sgrad4}) of the SF model will be shown at the end of this section. 
\begin{figure}[b]
\scalebox{0.6}[0.6]{\includegraphics{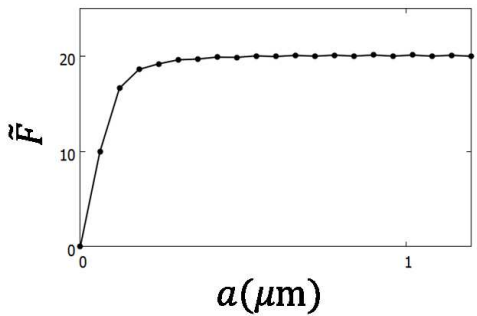}}
\caption{(Color online) Free energy difference $F_{eq}(a)-F_{eq}(0)$ computed with {\it no} eq.(\ref{eq:FLgrad2}) in the case of the A phase at 2.08(mK) and at 30 bar, which becomes $a$-independent with increasing $a$. 
} 
\label{fig.8}
\end{figure}

First, we explain the obtained results of the order parameter's spatial profile and $F_{eq}(a) - F_{eq}(0)$ in the bulk A phase, where the ${\bf l}$-vector is oriented to the ${\hat z}$-axis far from vortex cores, e.g., due to the slab geometry (see sec.I). To enable us to study a wider temperature range of the A phase, we focus here on the results at 30 bar. In this case, the variational GL equations are obtained from the sum of the bulk free energy terms used to obtain the phase diagram and the gradient energy terms (\ref{eq:gradquad}) with eq.(\ref{eq:WCc}), and eq.(\ref{eq:FLgrad4}). The resulting profile of each nonvanishing component of the order parameter is shown in Fig.6 (a) and (b), where the order parameter is represented by $A_{a,b}$ ($a$, $b = \pm$), defined in sec.VI, rather than $A_{\mu,i}$. As the left figures of Fig.6(a) show, a HQV pair is represented as a pair of integer vortices of $A_{++}$ and $A_{-+}$. In contrast to the expectation in London limit, however, $F_{eq}(a)$ increases with increasing $a$ at least near the vortex cores reflecting that $r_A > 1$ due to the SC corrections to the $\beta_j$ parameters mentioned in sec.IV (see Fig.7). In addition to this, as the right figures of Fig.6 (a) show, the components $A_{\pm-}$ with the opposite orbital chirality appear in a range around the vortex cores. It can be seen that, even if setting the boundary condition with ${\bf l} \parallel  +{\hat z}$, the order parameter component with ${\bf l} \parallel -{\hat z}$ tends to appear near the vortex cores. This feature, seen also in numerical results on the ordinary GL equations \cite{Nakahara} with no O($|\Delta|^4$) gradient term, seems to be another origin of elevating the energy of the HQV pair in the bulk A-phase, since the spatial region in which the components with the opposite chirality are nonvanishing becomes wider on approaching the superfluid transition temperature. As is seen below, such an excitation of unfavorable components of $A_{\mu,i}$ does not occur in the case of the polar phase which, as is seen later, seems to be a stage on which the HQV appears more easily. 

\begin{figure}[t]
\scalebox{0.35}[0.35]{\includegraphics{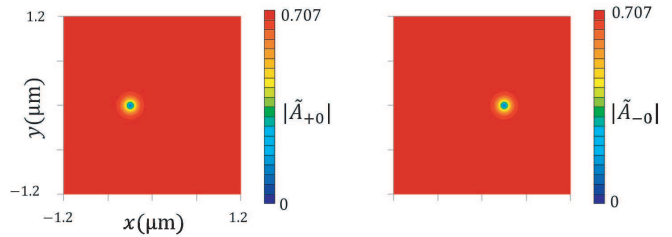}}
\caption{(Color online) Example of spatial profiles of the nonvanishing components $A_{\pm 0}$ of the order parameter mapped on the $x$-$y$ plane in the case with a HQV pair in the polar phase in a medium modelling \cite{AI06} an anisotropic aerogel. This figure has been obtained consistently with derivation of Fig.10(b).
} 
\label{fig.9}
\end{figure}
Figures 7 (a) and (b) show the corresponding free energy difference normalized properly (see the figure caption). Just below $T_c$ ($=2.44$ (mK)), i.e., at $T=2.43$ (mK), the contribution of the O($|\Delta|^4$) gradient term (\ref{eq:FLgrad4}) is too small to make a reduction of $F_{eq}(a)$ at larger $a$ visible, and, reflecting the cost of the free energy near the vortex cores due to the SC corrections and the mixing of the $A_{\mu,i}$ components with the opposite orbital chirality, the energy of a HQV pair is never lowered (see Fig.7 (a)). It implies that the HQV is not realized even as a metastable state very close to $T_c$. 
\begin{figure}[b]
\scalebox{0.6}[0.6]{\includegraphics{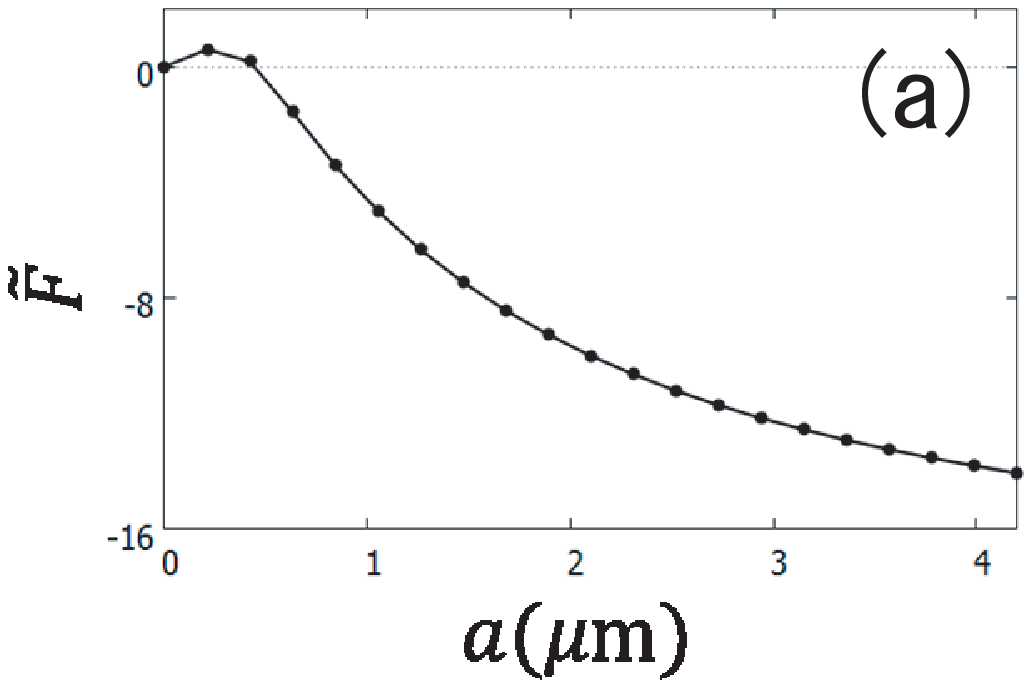}}
\scalebox{0.6}[0.6]{\includegraphics{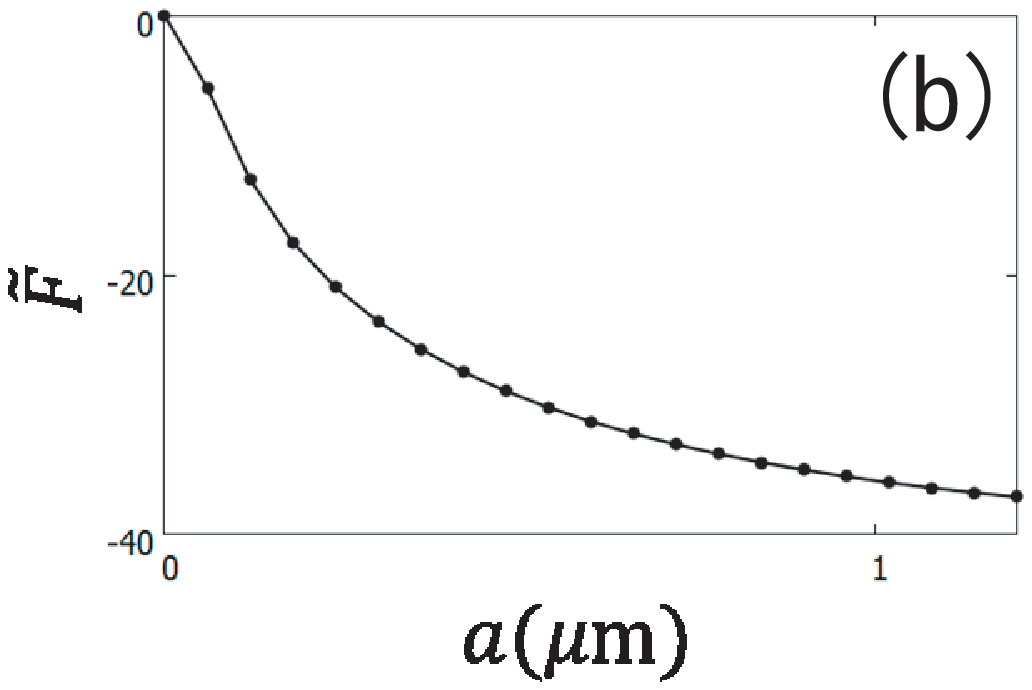}}
\caption{(Color online) Normalized free energy 
${\tilde F} \equiv \beta_{pol}(F_{eq}(a) - F_{eq}(0))/\alpha^2$ in the polar phase at two different temperatures and at 9 bar in Fig.5 (b)
. The figures (a) and (b) are the results at 1.73 (m) (just below the superfluid transition point) and 1.36 (mK) (just above the polar to A transition point), respectively. 
} 
\label{fig.10}
\end{figure}

For comparison, we show in Fig.8 the corresponding result of the free energy difference obtained at 2.08(mK) just above the AB transition temperature at 30 bar {\it without} the new gradient energy term (\ref{eq:FLgrad2}). Since the WC ($|\Delta|^4$) term (\ref{eq:Gorboxgrad}) is taken into account in obtaining Fig.8, the results in this figure that $F_{eq}(a) > F_{eq}(0)$, and that $F_{eq}$ is $a$-independent at large $a$ imply that the HQV-pair should not be realized at all in the conventional WC approximation with no term corresponding to the FL correction to the gradient energy. In contrast, the similar result in Fig.7(a) simply implies that the O($|\Delta|^4$) gradient term (\ref{eq:FLgrad2}) is ineffective in the close vicinity of $T_c$. 

\begin{figure}[t]
\scalebox{0.6}[0.6]{\includegraphics{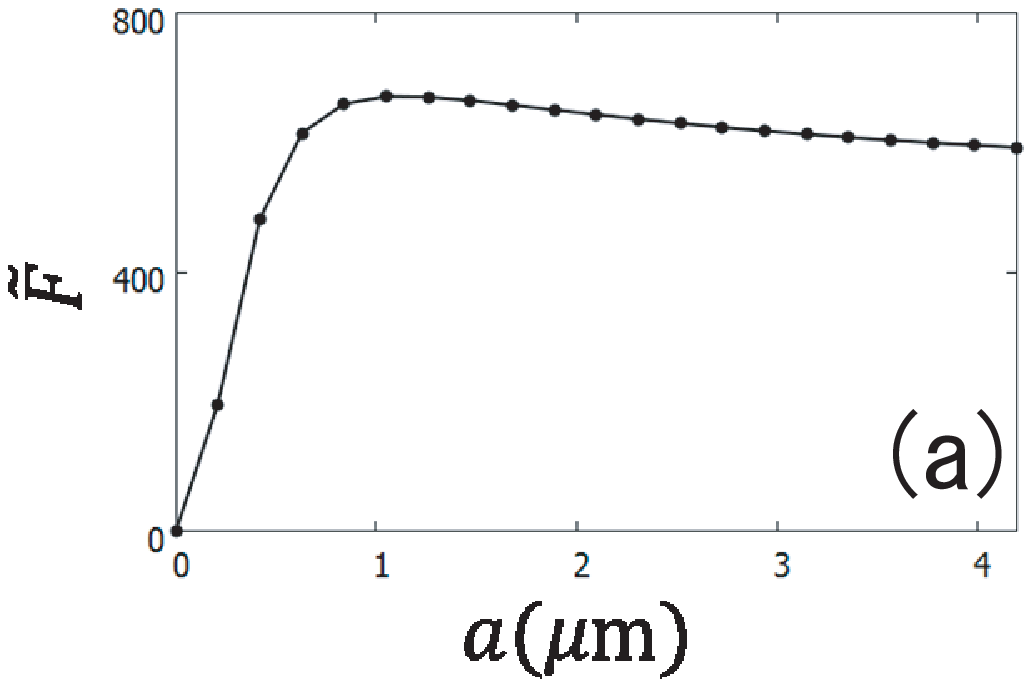}}
\scalebox{0.6}[0.6]{\includegraphics{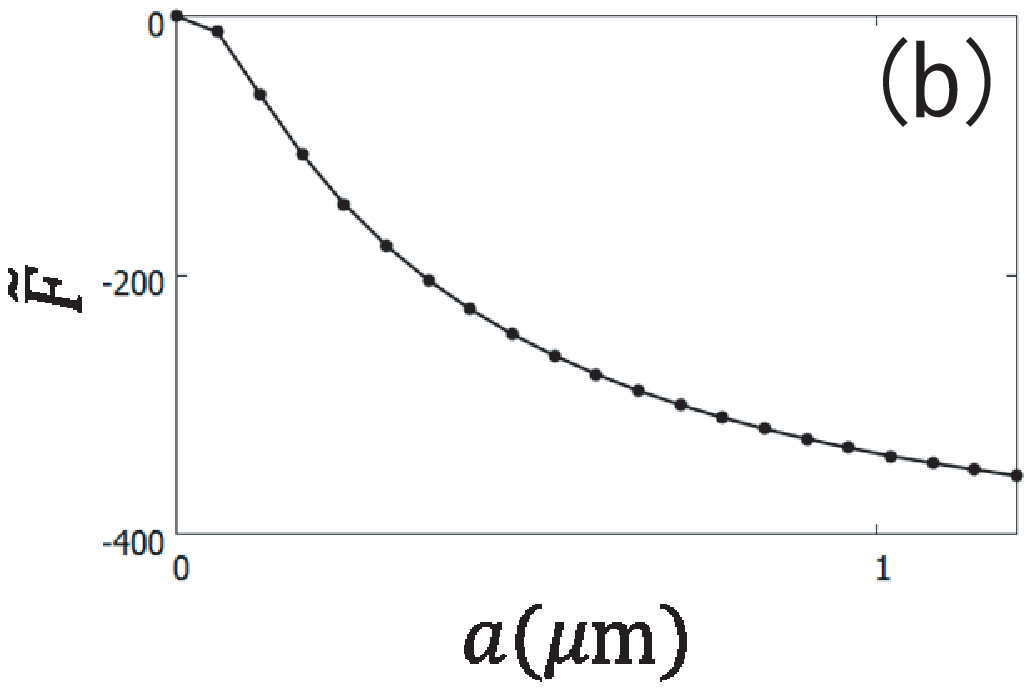}}
\scalebox{0.6}[0.6]{\includegraphics{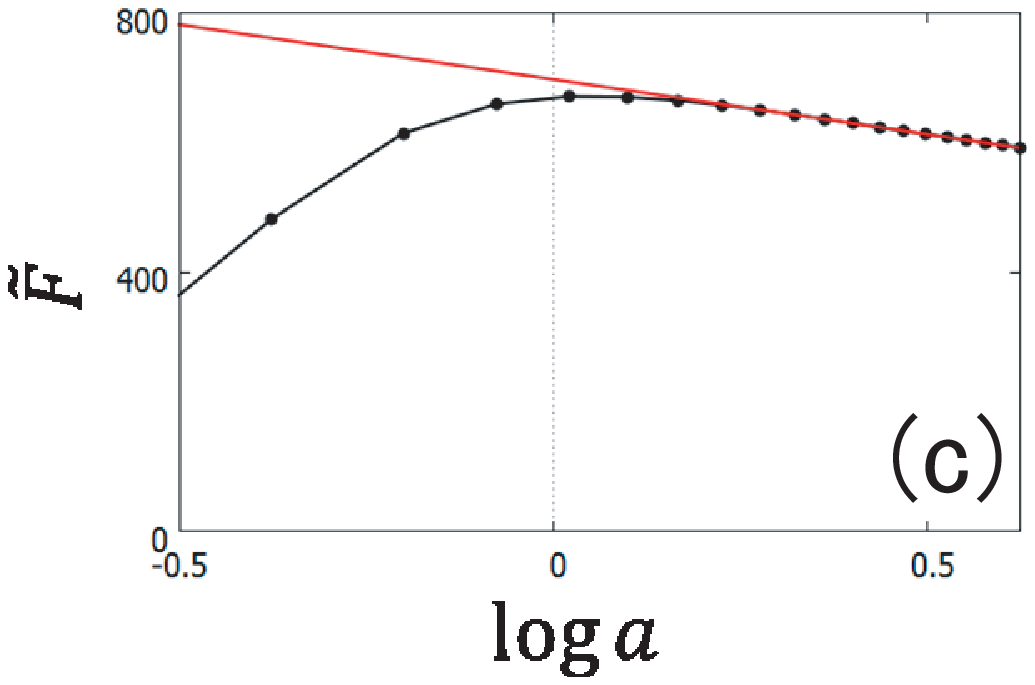}}
\caption{(Color online) The figures (a) and (b), corresponding to Fig.7 (a) and (b), respectively, are results corresponding to Fig.7 obtained using eq.(\ref{eq:Sgrad4}) in place of eq.(\ref{eq:FLgrad4}) and the parameter value ${\overline I}=0.724$. The figure (c) is the log($a$)-plot of the figure (a). 
} 
\label{fig.11}
\end{figure}

\begin{figure}[t]
\scalebox{0.6}[0.6]{\includegraphics{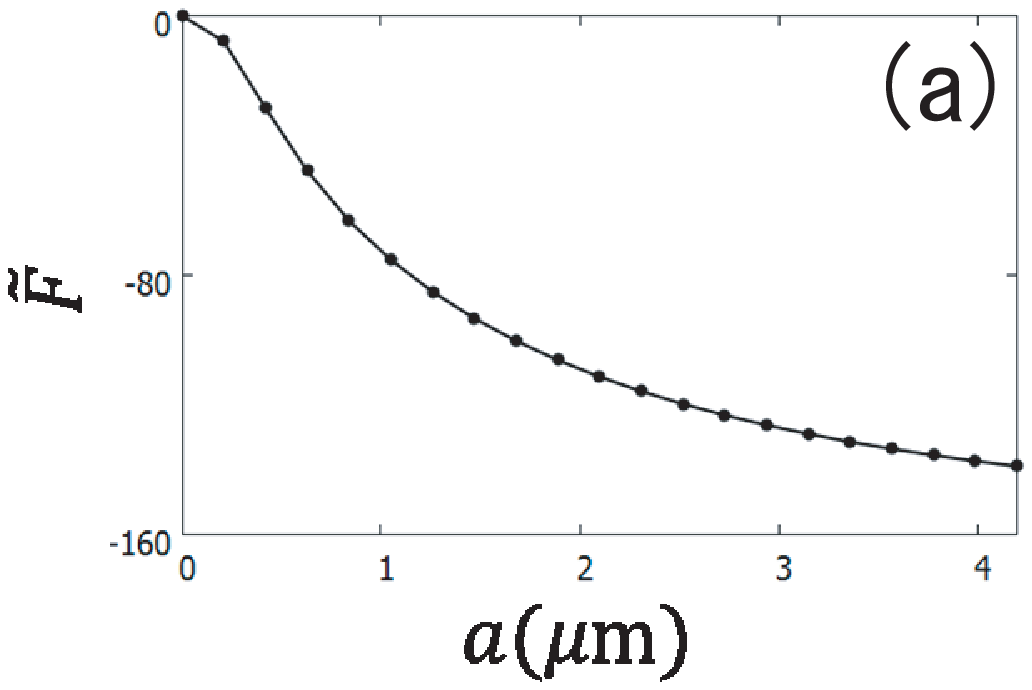}}
\scalebox{0.6}[0.6]{\includegraphics{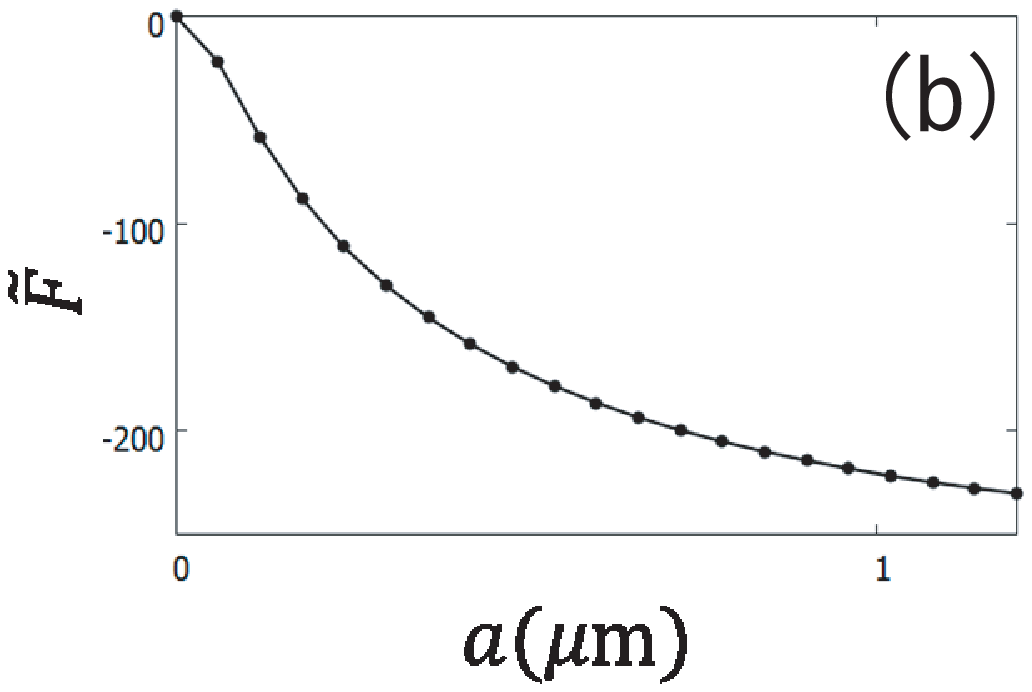}}
\caption{(Color online) Results corresponding to Fig.10 obtained using eq.(\ref{eq:Sgrad4}) in place of eq.(\ref{eq:FLgrad4}) and the parameter value ${\overline I}=0.724$. The figures (a) and (b) correspond to Fig.10 (a) and (b), respectively. 
} 
\label{fig.12}
\end{figure}

On the other hand, as is seen in Fig.7 (b), the free energy $F_{eq}(a)$ computed with the term (\ref{eq:FLgrad2}) decreases with increasing $a$, like in eq.(\ref{eq:FL2}), moderately below $T_c$ (see also Fig.11(c)). It strongly suggests that a HQV pair with a size $a$ of ten microns or so becomes stable. Nevertheless, the feature seen over a temperature range below $T_c$ that $F_{eq}(a) > F_{eq}(0)$ for smaller $a$ values should be noted because it may facilitate the situation in which PVs coexist with HQV-pairs in particular on a cooling upon rotation \cite{comexp}. 
Therefore, broadly speaking, the present results applicable to the bulk A phase lying only at higher pressures above 20 bar suggest that, at higher temperatures, the PVs tend to coexist with HQVs due to the importance of the SC correction to the condensation energy, while only the HQVs may be stabilized at lower temperatures close to the AB phase boundary. This feature seems to be consistent with a recent experimental result \cite{Ishikawa}, in which the measurement has been performed just near $T_c$, and coexistence of PVS and HQVs has inevitably 
occurred. 

\begin{figure}[b]
\scalebox{0.6}[0.6]{\includegraphics{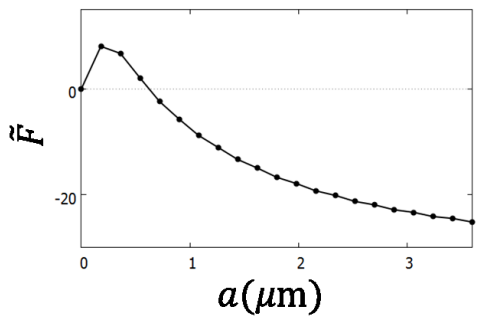}}
\caption{(Color online) Normalized free energy 
${\tilde F} \equiv \beta_{pol}(F_{eq}(a) - F_{eq}(0))/\alpha^2$ in the polar phase close to $T_c$ (2.35 (mK)) at 30 bar in Fig.5 (b) obtained using eq.(\ref{eq:Sgrad4}), i.e., based on the SF model. 
} 
\label{fig.13}
\end{figure}
Next, we turn to the case in a strongly anisotropic aerogel where the polar phase is realized. By using eqs.(\ref{eq:GLfb}), (4), (\ref{eq:SCbeta0}), (\ref{eq:gradquad}), (\ref{eq:WCgradani}), (\ref{eq:SCgradani}), (\ref{eq:SCbetavc}), (\ref{eq:GLfbani}), and (\ref{eq:FLgrad4}) and solving the resulting variational GL equations with the parameter values used in obtaining the phase diagram Fig.5 (b), numerical solutions of one HQV pair are obtained. As one example, we show here the obtained results at 9 bar, since the NMR data in Ref.\cite{Autti16} have been taken at a low pressure, 7 bar. In this case, in contrast to the bulk A phase case, the temperature width of the polar phase is wider in such lower pressures where the SC correction indicated in eq.(\ref{eq:SCbeta0}) tending to destabilize the HQV is weaker. Thus, the situation in which the HQVs are realized is expected to be prepared more easily in the polar phase in an aerogel. In fact, the obtained profiles of the order parameters $A_{\mu,i}$ are consistent with this expectation: The components $A_{\pm 0}$ shown in Fig.9 are the only nonvanishing components of $A_{ab}$ obtained in the polar phase under the outer boundary condition, eq.(\ref{eq:phihq}). So, the HQV in the polar phase has a much simpler structure than that in the bulk A phase. The resulting pair size dependences of the free energy close to the normal to polar superfluid transition at $1.74$ (mK) and at the low temperature, 1.36 (mK), are shown in Fig.10. In contrast to the bulk A phase, even in Fig.10 (a) taken close to $T_c$, the small $a$ range in which $F_{eq}(a) > F_{eq}(0)$ is extremely narrow, and $F_{eq}(a)$ decreases with increasing $a$ for almost all $a$ values. Thus, even near $T_c$, a HQV-pair is expected to be more stable in the polar phase in aerogels at least at such low pressures. 
This HQV's stability seems to be a consequence of the two features: One is the fact that the measure of the SC effect $r_{pol}$, defined in sec.VII, is low and close to unity reflecting a cancellation between the SC effect and the impurity scattering effect in the $\beta_j$ parameters (see sec.VII), and the other is the simpler order parameter profiles in real space shown in Fig.9. 

Our results shown above were obtained by using the gradient energy (\ref{eq:FLgrad2}) derived based on the conventional Fermi liquid theory as the additional term stabilizing the HQV pairs. Before ending this section, the corresponding results obtained by replacing eq.(\ref{eq:FLgrad4}) with the interaction-induced gradient term (\ref{eq:Sgrad4}) in the SF approach \cite{BSA} will be shown. Figure 11 and 12 are the corresponding results to those of Figs.7 and 10. Quantitatively, the quantum SF model of the repulsive interaction between the quasiparticles seems to stabilize the HQV pair further. Nevertheless, our conclusions on the HQV-pair's stability in the bulk A phase and the polar phase in the anisotropic aerogels are qualitatively the same irrespective of which of eqs.(\ref{eq:Sgrad4}) and (\ref{eq:FLgrad4}) is used. 

As in the bulk A phase, the range of the $a$ values in which $F_{eq}(a) > F_{eq}(0)$ becomes visible even in the polar phase at higher pressures as a result of the SC effect enhanced with increasing the pressure. In Fig.13 taken just below $T_c$ and at 30 bar, this feature is clearly seen. Nevertheless, such a range is too narrow to make the HQV pairs unstable. 
Based on these results obtained from the impurity scattering model \cite{AI06} on the superfluid $^3$He in anisotropic aerogel, it is believed that the emergence of the HQVs in the polar phase in anisotropic aerogels \cite{Dmitriev15,Autti16} is not a metastable event assisted by the pinning to the aerogel structures but an intrinsic event. 

\section{Summary and Discussion} 

As mentioned in sec.I, there has been a gap so far on theoretical understanding of the HQVs in superfluid $^3$He between the London limit and the conventional GL theory. For instance, in the A phase with ${\bf l}$-vector perpendicular to the plane in a slab geometry, the treatment in the London limit predicts that, as a consequence of the Fermi liquid (FL) correction to the gradient terms, a HQV pair is more stable than a single PV, while the conventional GL free energy based on the familiar weak-coupling approximation does not include such a gradient term leading to the HQV-pair's stability. Clearly, the use of the {\it conventional} GL free energy \cite{Machida,Nakahara} is not appropriate for studying the HQVs' stability. To bridge this gap on theoretical descriptions of a HQV pair, we have microscopically examined the gradient energy terms in the GL free energy. Depending on the way of describing the SC effect on the bulk free energy \cite{BSA,SS}, the two approaches for describing the effects of the repulsive interaction between the quasiparticles on the GL gradient term contributing to the HQVs' stability can be considered. One is the Fermi liquid model, and the other is the spin fluctuation model. By deriving the corresponding interaction-induced gradient energy term in the two approaches, we have performed numerical computations on the resulting extended GL free energy and have reached the conclusion that, in the bulk A phase in a slab geometry, HQV pairs may be stabilized without coexistence with PVs far below $T_c$, while, in the case of the polar phase realized in anisotropic aerogels, a HQV pair is certainly stable even close to $T_c$ in particular at low pressures. 

We note that the stability of a HQV pair in the polar phase is unexpected in the following sense: As mentioned in the sentences below eq.(\ref{eq:FLgradL}) and (\ref{eq:difK2}), the interaction-induced gradient term stabilizing the HQVs in the polar phase is {\it smaller} than that in the bulk A phase, and thus that, once one is based on the London limit and takes account only of the gradient terms, it is difficult to understand why the HQVs have been more clearly realized in the polar phase. In fact, the importance of the vortex pinning via the aerogel structure for the emergence of the HQVs has been stressed in Ref.\cite{Autti16}. On the other hand, our results obtained by taking account of both the condensation energy and the gradient energy suggest that the stability of HQVs in the polar phase in aerogels can be understood without invoking the pinning effect. At the present stage, it is unclear to us to what extent the vortex pinning effect due to the aerogel structure assists the stability of HQVs realized in experiments \cite{Autti16}.

Emergence of a HQV pair in the bulk A phase is limited to some extent because the SC contribution tending to destabilize the HQVs is more effective at higher pressures where the bulk A phase is realized. It is speculated that the coexistence of HQVs and PVs found in the bulk A phase \cite{Ishikawa} is due not to the presence of a texture of the ${\bf l}$-vector in a slab geometry with a large film thickness but to the pressure-induced SC effect. In contrast, the fact that the polar phase region in anisotropic aerogels is wider at lower pressures \cite{Dmitriev15,RI15} seems to have assisted realization of HQVs. 

It should be stressed that the situation in which the A phase is realized at {\it low} pressures in a quasi two-dimensional geometry where the ${\bf l}$-vector is fixed perpendicularly to the plane may be another candidate for realization of the HQV. Such a situation is seen, e.g., in Fig.7c in Ref.\cite{Bil18} and Ref.\cite{Saunders}. 

We have not considered the HQVs in the A phase \cite{Dmitriev15,AI06} occurring at lower temperatures than the polar phase in the anisotropic aerogels in the present work, because talking account of effects of the texture of the ${\bf l}$-vector which is inevitably present in this phase is beyond the scope of the present work. But, this A phase is also one of the A phases occurring at lower pressures commented in the last paragraph. The measurements in Ref.\cite{Autti16,Kibble} have shown that the HQVs seen in the polar phase survive in this A phase where the Majorana fermions should exist as their core state \cite{Ivanov}. Clearly, it is an intriguing subject to examine effects of the disorder-induced texture of the ${\bf l}$-vector on the HQV-stability. In relation to this, the possibility of a HQV pair in the polar phase upon rotation around an axis perpendicular to the anisotropy axis of the aerogel should be considered, because the HQV in this case may be movable in contrast to the case upon rotation parallel to the anisotropy axis \cite{Autti16}. 

The present theory taking account of an additional gradient term in the GL free energy might play a significant role in describing the vortices in the B phase \cite{Th86,SV2} because the gradient energy is more important near the vortex cores. According to our preliminary results, the O($|\Delta|^4$) gradient term, eq.(\ref{eq:FLgrad4}), stabilizes the nonaxisymmetriv vortex with the core consisting of a HQV-pair : The FL-based gradient term (\ref{eq:FLgrad2}) assists the stability of this vortex, while the weak-coupling higher-order gradient term (\ref{eq:Gorboxgrad}) tends to destabilize it. These subjects should be considered further in future works.

The present research was supported by JSPS KAKENHI [Grant No. 16K05444].

\section{Appendix} 

In this Appendix, expressions which were omitted in the text but are to be used for numerical calculations will be presented. 

\begin{figure}[b]
\scalebox{0.35}[0.35]{\includegraphics{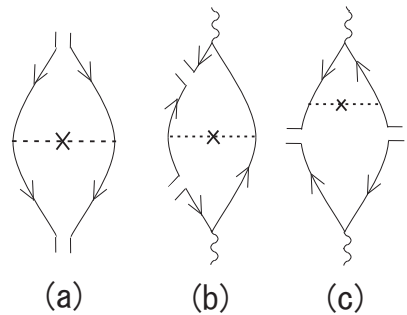}}
\caption{(Color online) (a) : Lowest order diagram in the anisotropy $\delta_u$ leading to eq.(\ref{eq:WCgradani}). (b) and (c) : Quantitatively dominant diagrams expressing $\delta \chi_{\alpha \alpha}$ in the lowest order in $\delta_u$. The wavy line denotes the SF propagator in the normal state. 
} 
\label{AppH}
\end{figure}

In the WC approximation, the coefficients in eq.(\ref{eq:gradquad}) are given by 
\begin{equation}
K_1^{({\rm wc})} = K_2^{({\rm wc})} = \frac{N(0)}{60} \biggl(\frac{v_{\rm F}}{2 \pi T} \biggr)^2 |\psi^{(2)}(y)|,
\label{eq:WCc}
\end{equation}
where $y=(1  + 1/(2 \pi \tau T))/2$, and $\tau^{-1}$ denotes the relaxation rate of a quasiparticle via the elastic impurity scattering with the aerogel structure. 
As far as the system is isotropic, the relation $K_1=K_2$ remains valid even if the SF-induced corrections to eq.(\ref{eq:gradquad}) which occurs from eq.(\ref{eq:BSA}) are included. Consistently with the neglect of the contributions of Fig.3(c) to the quartic bulk energy terms, these corrections to eq.(\ref{eq:gradquad}) may be neglected for simplicity so that $K_j$ in eq.(\ref{eq:gradquad}) will be identified with $K_j^{({\rm wc})}$ in our numerical analysis. 
In aerogels, the anisotropy-induced correction to the WC gradient term of O($|\Delta|^2$) resulting from the diagram of Fig.\ref{AppH} (a), up to O($\delta_u$), takes the form 
\begin{eqnarray}
f_{grad,ani}^{(wc)} &=& - \frac{N(0)}{1080} \frac{\delta_u}{2 \pi T \tau} \biggl(\frac{v_{\rm F}}{2 \pi T} \biggr)^2 \psi^{(3)}(y) [ 2 \partial_j A_{\mu,z}^* \partial_j A_{\mu,z} \nonumber \\
&-& \partial_i A_{\mu,i}^* \partial_j A_{\mu,j} + (\partial_j A_{\mu,j}^* \partial_z A_{\mu,z} + {\rm c.c.} ) ], 
\label{eq:WCgradani} 
\end{eqnarray}
where $\delta_u$ is the parameter measuring the anisotropy strength introduced in Ref.2. 

On the other hand, the main correction terms due to the SF to the anisotropic gradient term of O($|\Delta|^2$), which result from Figs.\ref{AppH} (b) and (c), are expressed by  
\begin{widetext}
\begin{eqnarray}
f_{grad,ani}^{(sc)} &=& \frac{\pi^2 N(0)}{3360} (\pi {\overline I})^2 \frac{\delta_u}{2 \pi T \tau} \biggl(\frac{v_{\rm F}}{2 \pi T} \biggr)^2 \biggl[ \biggl( L_1+ \frac{3}{2} L_2 \biggr) \partial_i A_{\mu,j} \partial_i A^*_{\mu,j} + \biggl(16 L_1 + \frac{27}{16} L_2 \biggr) \partial_i A_{\mu,z} \partial_i A^*_{\mu,z} \nonumber \\ 
&+& \biggl(-5 L_1 + \frac{27}{16} L_2 \biggr) (\nabla \cdot A_\mu) (\nabla \cdot A^*_{\mu}) + \biggl(2 L_1 + \frac{3}{4} L_2 \biggr) \partial_z A_{\mu,i} \partial_z A^*_{\mu,i} \nonumber \\
&+& \biggl(11 L_1 + \frac{135}{32} L_2 \biggr) ((\nabla \cdot A_\mu) \partial_z A_{\mu,z}^* + {\rm c.c.} ) \biggr],  
\label{eq:SCgradani}
\end{eqnarray}
\end{widetext}
where 
\begin{eqnarray}
L_1 &=& \psi^{(3)}(y) \sum_m D_1(|m|), \nonumber \\ 
L_2 &=& [\psi^{(3)}(y) - \psi^{(3)}(y+|m|)] D_1(|m|), 
\label{anic}
\end{eqnarray}
and 
\begin{equation}
D_1(|m|) = \frac{T}{8 \pi^2 E_{\rm F}} \int_0^\infty d{\overline q} \biggl(1 - {\overline I} 
+ \frac{{\overline I}}{3} {\overline q}^2 + \frac{\pi^2 T}{4|{\overline q}|E_{\rm F}}|m| \biggr)^{-1}. 
\label{d1m}
\end{equation}

Next, the terms to be added in the bulk free energy (i.e., except the gradient terms) in the case with impurity scattering processes will be briefly 
explained. 
In considering the vortices in $^3$He in aerogels, additional contributions to $\beta_j$ arising from the impurity-scattering effects due to the aerogel need to be taken into account. They were obtained in Ref.\cite{RI15}, and, when only the unitary pairing states are assumed to be realized, their main contributions are given by
\begin{widetext}
\begin{eqnarray}
\beta_{2,{\rm vc}}^{({\rm sc})} &=& 30 \frac{(\pi {\overline I})^2}{2 \pi \tau T} {\overline \beta}_0 \sum_m D_1(|m|) \biggl[\biggl(\frac{2}{15} + \frac{\pi^2}{80} \biggr) \psi^{(3)}(y) + \frac{\pi^2}{40} (\psi^{(3)}(y) - \psi^{(3)}(y+|m|)) \biggr], \nonumber \\
\beta_{3,{\rm vc}}^{({\rm sc})} &=& \beta_{4,{\rm vc}}^{({\rm sc})} = 30 \frac{(\pi {\overline I})^2}{2 \pi \tau T} {\overline \beta}_0 \sum_m D_1(|m|) \biggl[\biggl(\frac{2}{15} + \frac{\pi^2}{480} \biggr) \psi^{(3)}(y) + \frac{\pi^2}{240} (\psi^{(3)}(y) - \psi^{(3)}(y+|m|)) \biggr], 
\label{eq:SCbetavc}
\end{eqnarray} 
\end{widetext} 
while the corresponding $\beta_{1,{\rm vc}}^{({\rm sc})}$ and $\beta_{5,{\rm vc}}^{({\rm sc})}$ are zero. 

In evaluating the condensation energy of the polar phase in an anisotropic aerogel, anisotropy-induced terms in the GL free energy need to be taken into account in addition to the six terms in eq.(\ref{eq:GLfb}). In the case of superfluid $^3$He in a globally anisotropic aerogel, such additional terms to the GL free energy take the form 
\begin{widetext}
\begin{eqnarray}
F_{bulk, ani} &=& \int d^3r \biggl[(\alpha^{({\rm sc})}|_{\rm se} + \alpha^{({\rm sc})}|_{\rm vc}) A_{\mu j}^* A_{\mu j} + (\alpha_z^{({\rm wc})} + \alpha_z^{({\rm sc})}|_{\rm se} + \alpha_z^{({\rm sc})}|_{\rm vc}) A_{\mu z}^* A_{\mu z} + [ \beta_{1z}^{({\rm wc})} A_{\mu i}A_{\mu i}A^*_{\mu z}A^*_{\mu z} \nonumber \\ 
&+& \beta^{({\rm wc})}_{2z} A^*_{\mu i}A_{\mu i}A^*_{\nu z}A_{\nu z} 
+ \beta^{({\rm wc})}_{3z} A_{\mu i}A_{\nu i} A^*_{\mu z}A^*_{\nu z} + \beta^{({\rm wc})}_{4z} A^*_{\mu i} A_{\nu i} A^*_{\nu z} A_{\mu z} + \beta^{({\rm wc})}_5 A_{\mu i}^*A_{\nu i}A_{\mu z}^*A_{\nu z} 
+ {\rm c.c.}] \biggr] 
\label{eq:GLfbani}
\end{eqnarray}
\end{widetext}
in the notation in Refs.\cite{AI06,RI15}. In addition to $\beta^{({\rm wc})}_{jz}$, anisotropy-induced corrections also appear in $\beta^{({\rm wc})}_j$ defined in eq.(\ref{eq:GLfb}). Their detailed expressions were given in Ref.\cite{RI15} together with the coefficients of the quadratic terms in eq.(\ref{eq:GLfbani}). By applying them to the ratio $r_{pol}$ defined in sec.VII, we obtain $r_{pol}$ given in eq.(\ref{eq:ratio4}), where 
\begin{equation}
c_r= \frac{1}{2 \pi \tau T} \biggl|\frac{\psi^{(3)}(y)}{\psi^{(2)}(y)} \biggr| \biggl(\frac{-331\delta_u - 35}{378} \biggr). 
\label{eq:ratio3}
\end{equation}

Finally, we will explain how to derive the gradient terms of O($|\Delta|^4$). 
The WC contribution to the gradient term of O($|\Delta|^4$) follows from the full expression of Fig.1 
\begin{widetext}
\begin{eqnarray}
F_4^{(wc)} &=& \frac{\beta^{-1}}{2} \sum_\varepsilon \sum_{{\bf k}_1, \cdot \cdot \cdot {\bf k}_4} \delta_{{\bf k}_1+{\bf k}_3, {\bf k}_2+{\bf k}_4} \int \frac{d^3{\bf p}}{(2 \pi)^3} {\rm Tr}(\sigma_\mu \sigma_\nu \sigma_\rho \sigma_\lambda) {\cal G}_p(\varepsilon) {\cal G}_{-p+k_1}(-\varepsilon) {\cal G}_{p+k_2-k_1}(\varepsilon) {\cal G}_{-p+k_4}(-\varepsilon) \nonumber \\ 
&\times& D_\mu({\bf p}; {\bf k}_1) D^*_\nu({\bf p}; {\bf k}_2) D_\rho({\bf p}; {\bf k}_3) D^*_\lambda({\bf p}; {\bf k}_4). 
\label{eq:Gorbox}
\end{eqnarray}
\end{widetext}
Alternatively, this expression may be expressed in the form of eq.(18), where 
\begin{eqnarray}
f({\bf v}\cdot{\bf k}_j) &=& \frac{\beta^{-1}}{2} N(0) \sum_{\varepsilon} \int d\xi \, {\cal G}_p(\varepsilon) \, {\cal G}_{-p+k_1}(-\varepsilon) \nonumber \\ 
&\times& {\cal G}_{p+k_2-k_1}(\varepsilon) {\cal G}_{-p+k_4}(-\varepsilon) 
\label{eq:fvk}
\end{eqnarray}
with the single particle kinetic energy measured from the Fermi energy $\xi$. 
By picking the quadratic terms in $k_i$ up from the product of the Green's functions and rewriting the O($k^2$) terms in eq.(\ref{eq:Gorbox}) 
in the real space 
representation, 
the contribution to the gradient terms of O($|\Delta|^4$) in the WC approximation becomes 
\begin{widetext}
\begin{eqnarray}
F_{grad4}^{(wc)} &=& N(0) \frac{v_{\rm F}^2}{384 (2 \pi T)^4} \psi^{(4)}(y) {\rm Tr}(\sigma_\mu \sigma_\nu \sigma_\rho \sigma_\lambda) \int d^3{\bf r} \langle {\hat p}_i {\hat p}_j [D_\mu({\bf p}) \partial_i D^*({\bf p})_\nu D_\rho({\bf p}) \partial_j D^*_\lambda({\bf p}) \nonumber \\ 
&+& \partial_i D_\mu({\bf p}) D^*_\nu({\bf p}) \partial_j D_\rho({\bf p}) D^*_\lambda({\bf p}) 
+ \frac{3}{2} (D_\mu({\bf p}) \partial_i D^*_\nu({\bf p}) \partial_j D_\rho({\bf p}) D^*_\lambda({\bf p}) + \partial_i D_\mu({\bf p}) D^*_\nu({\bf p}) D_\rho({\bf p}) \partial_j D^*_\lambda({\bf p}) \nonumber \\
&+& \partial_i D_\mu({\bf p}) \partial_j D^*_\nu({\bf p}) D_\rho({\bf p}) D^*_\lambda({\bf p}) + D_\mu({\bf p}) D^*_\nu({\bf p}) \partial_i D_\rho({\bf p}) \partial_j D^*_\lambda({\bf p})) ], 
\label{eq:Gorboxgrad}
\end{eqnarray}
\end{widetext}
where $D_\mu({\bf p})$ is the Fourier transform of $D_\mu({\bf p}; {\bf k})$. 
The final expression of eq.(\ref{eq:Gorboxgrad}) will be summarized below together with the corresponding one of Fig.3(c) or Fig.4. 

The expression of Fig.3 (c) is given by the first line of eq.(\ref{eq:BSA}) 
with 
\begin{widetext}
\begin{eqnarray}
\delta \chi_{\alpha \alpha}({\bf q}) &=& - \frac{\beta^{-1}}{2} {\rm Tr}(\sigma_\mu \sigma_\nu \sigma_\alpha \sigma_\rho \sigma_\lambda \sigma_\alpha) \sum_{k_1 \cdot \cdot \cdot k_4} \delta_{k_1+k_3, k_2+k_4} \int \frac{d^3{\bf p}}{(2 \pi)^3} \sum_\varepsilon {\cal G}_{\bf p_-}(\varepsilon) {\cal G}_{-\bf p_-+k_1}(-\varepsilon) {\cal G}_{\bf p_-+k_2-k_1}(\varepsilon) \nonumber \\
&\times& {\cal G}_{\bf p_++k_2-k_1}(\varepsilon+\Omega) {\cal G}_{-\bf p_++k_4}(-\varepsilon-\Omega) {\cal G}_{\bf p_+}(\varepsilon+\Omega) D_\mu({\bf p}; {\bf k}_1) D^*_\nu({\bf p}; {\bf k}_2) D_\rho({\bf p}; {\bf k}_3) D^*_\lambda({\bf p}; {\bf k}_4). 
\label{eq:delchi}
\end{eqnarray}

The corresponding gradient free energy term is given by 

\begin{equation}
F_{Sgrad 4} = - \frac{{\overline I}^2}{2 N(0)} T \sum_\Omega \int_{\bf q} \frac{1}{1 - I \chi_N({\bf q}, \Omega)} \delta \chi''_{\alpha, \alpha}({\bf q}, \Omega),
\label{eq:SCgrad4}
\end{equation}
where $\delta \chi''$ is the gradient term of $\delta \chi$, and, if, for simplicity, focusing on its $\Omega=0$ term arising from the thermal SF, it is expressed as 
\begin{eqnarray}
\delta \chi''_{\alpha, \alpha}({\bf q}, 0) 
&=& - \frac{\pi}{16} \beta^{-1} N(0) v_{\rm F}^2 {\rm Tr}(\sigma_\mu \sigma_\nu \sigma_\alpha \sigma_\rho \sigma_\lambda \sigma_\alpha) 
\sum_\varepsilon \biggl\langle \frac{{\hat p}_i {\hat p}_j}{|\varepsilon|^5 [({\bf v}\cdot{\bf q})^2 + 4|\varepsilon|^2]} \int d^3{\bf r} \biggl
[3 \biggl(D_\mu({\bf p}) \partial_i D^*_\nu({\bf p}) \partial_j D_\rho({\bf p}) D^*_\lambda({\bf p}) \nonumber \\
&+& \partial_i D_\mu({\bf p}) D^*_\nu({\bf p}) D_\rho({\bf p}) \partial_j D^*_\lambda({\bf p}) + \partial_i D_\mu({\bf p}) \partial_j D^*_\nu({\bf p}) D_\rho({\bf p}) D^*_\lambda({\bf p}) + D_\mu({\bf p}) D^*_\nu({\bf p}) \partial_i D_\rho({\bf p}) \partial_j D^*_\lambda({\bf p}) \biggr) \nonumber \\ 
&+& (D_\mu({\bf p}) \partial_i D^*_\nu({\bf p}) D_\rho({\bf p}) \partial_j D^*_\lambda({\bf p})  
+ \partial_i D_\mu({\bf p}) D^*_\nu({\bf p}) \partial_j D_\rho({\bf p}) D^*_\lambda({\bf p})) \biggr] \biggr\rangle . 
\label{eq:delchidd}
\end{eqnarray}
Then, the gradient energy density of O($|\Delta|^4$) following from the sum of eqs.(\ref{eq:Gorboxgrad}) and eq.(\ref{eq:SCgrad4}) is given by

\begin{eqnarray}
f_{Sgrad4} &=& \frac{N(0)}{26880 (\pi T)^2} \biggl(\frac{v_{\rm F}}{2 \pi T} \biggr)^2 \biggl[u_{1+} \biggl( \, 
(\nabla\cdot A_\mu)(\nabla\cdot A_\lambda^*)A_{\mu i}^*A_{\lambda i} 
+A_{\lambda j}A_{\mu j}^*\partial_k A_{\lambda i}^*\partial_i A_{\mu k} 
+(\nabla\cdot A_{\mu})(\nabla\cdot A_{\mu}^*)A_{\lambda j}A_{\lambda j}^* \nonumber \\ 
&+& A_{\lambda j}A_{\lambda j}^*\partial_k A_{\mu i}^*\partial_i A_{\mu k} 
+(\nabla A_{\mu i})\cdot(\nabla A_{\mu i}^*)A_{\lambda j}A_{\lambda j}^* 
+ (\nabla A_{\mu i})\cdot(\nabla A_{\mu j}^*)(A_{\lambda i}A_{\lambda j}^*+A_{\lambda i}^*A_{\lambda j}) \nonumber \\
&+& (\nabla A_{\mu i})\cdot(\nabla A_{\lambda i}^*) A_{\mu j}^*A_{\lambda j} 
+ (\nabla A_{\mu i})\cdot(\nabla A_{\lambda j}^*)A_{\mu i}^*A_{\lambda j} 
+ (\nabla A_{\mu i})\cdot(\nabla A_{\lambda j}^*)A_{\mu j}^*A_{\lambda i} 
+ (A_\lambda\cdot\nabla)A_{\lambda i}^*(A_\mu^*\cdot\nabla)A_{\mu i} 
\nonumber \\ 
&+& (A_{\lambda}\cdot\nabla)A_{\mu i}^* (A_{\lambda}^*\cdot\nabla)A_{\mu i} 
+ (A_\lambda\cdot\nabla)A_{\mu i}(A_{\lambda}^*\cdot\nabla)A_{\mu i}^* 
+ (A_\lambda\cdot\nabla)A_{\mu i}(A_\mu^*\cdot\nabla)A_{\lambda i}^* 
 \nonumber \\
&+& [ (\nabla\cdot A_\mu)(A_{\mu i}^*(A_\lambda\cdot\nabla)A_{\lambda i}^* + A_{\lambda i}(A_{\mu}^*\cdot\nabla)A_{\lambda i}^*) 
+ (\nabla\cdot A_\mu)(A_{\lambda i}^*(A_\lambda\cdot\nabla)A_{\mu i}^* 
+ A_{\lambda i}(A_{\lambda}^*\cdot\nabla)A_{\mu i}^*) \nonumber \\
&+& A_{\mu i}^*((A_\lambda\cdot\nabla)A_\lambda^*\cdot\nabla)A_{\mu i} 
+ A_{\lambda i}((A_\mu^*\cdot\nabla)A_\lambda^*\cdot\nabla)A_{\mu i} 
+ A_{\lambda i}((A_{\lambda}^*\cdot\nabla)A_{\mu}^*\cdot\nabla)A_{\mu i}
+ A_{\lambda i}^*((A_{\lambda}\cdot\nabla)A_{\mu}^*\cdot\nabla)A_{\mu i} 
 + {\rm c.c.} ] \,\biggr) \nonumber \\
&+& u_{1-} \biggl((\nabla A_{\mu i})\cdot(\nabla A_{\lambda i}^*)A_{\mu j}A_{\lambda j}^* + (\nabla A_{\mu i})\cdot(\nabla A_{\lambda j}^*)(A_{\mu i}A_{\lambda j}^* 
+ A_{\mu j}A_{\lambda i}^*) 
+ (\nabla\cdot A_\mu)(\nabla\cdot A_\lambda^*)A_{\mu i}A_{\lambda i}^* \nonumber \\
&+& A_{\mu i}A_{\lambda i}^*\partial_k A_{\lambda j}^*\partial_j A_{\mu k} 
+ (A_{\lambda}^*\cdot\nabla)A_{\mu i}(A_\mu\cdot\nabla)A_{\lambda i}^*  + (A_{\mu}\cdot\nabla)A_{\mu i}(A_{\lambda}^*\cdot\nabla)A_{\lambda i}^* + [ (\nabla\cdot A_\mu)(A_{\lambda j}^*(A_\mu\cdot\nabla)A_{\lambda j}^* 
\nonumber \\ 
&+& A_{\mu j}(A_\lambda^*\cdot\nabla)A_{\lambda j}^*) 
+ A_{\mu i}((A_\lambda^*\cdot\nabla)A_\lambda^*\cdot\nabla)A_{\mu i} + A_{\lambda i}^*((A_\mu\cdot\nabla)A_\lambda^*\cdot\nabla)A_{\mu i} 
+ {\rm c.c.}] \biggr) \nonumber \\ 
&+& u_{2+} \biggl( (\nabla A_{\mu i}^*)\cdot(\nabla A_{\lambda j}^*)(A_{\mu i}A_{\lambda j} + A_{\mu j}A_{\lambda i}) + (\nabla A_{\mu i}^*)\cdot(\nabla A_{\lambda i}^*)A_{\mu j}A_{\lambda j} 
+ (A_{\mu}\cdot\nabla)A_{\mu i}^*(A_{\lambda}\cdot\nabla)A_{\lambda i}^* \nonumber \\
&+& (\nabla\cdot A_\mu^*)(\nabla\cdot A_\lambda^*)A_{\mu i}A_{\lambda i} 
+ (A_{\lambda}\cdot\nabla)A_{\mu i}^*(A_{\mu}\cdot\nabla)A_{\lambda i}^* 
+ A_{\mu i}A_{\lambda i}\partial_j A_{\mu k}^*\partial_k A_{\lambda j}^* 
+2[ (\nabla\cdot A_\mu^*)(A_{\lambda i}(A_\mu\cdot\nabla)A_{\lambda i}^* \nonumber \\
&+& A_{\mu i}(A_\lambda\cdot\nabla)A_{\lambda i}^*)+
A_{\lambda i}(((A_\mu\cdot\nabla)A_\mu^*\cdot\nabla)A_{\lambda i}^*+((A_\mu\cdot\nabla)A_\lambda^*\cdot\nabla)A_{\mu i}^*) ]
 + {\rm c.c.} \biggr) \nonumber \\
&+& u_{2-} \biggl( (\nabla A_{\mu i}^*)\cdot(\nabla A_{\mu j}^*)A_{\lambda i}A_{\lambda j} + (A_{\lambda}\cdot\nabla)A_{\mu i}^*(A_{\lambda}\cdot\nabla)A_{\mu i}^* \nonumber \\ 
&+& 2[(\nabla\cdot A_\mu^*)A_{\lambda i}(A_\lambda\cdot\nabla)A_{\mu i}^* 
+ A_{\lambda i}((A_\lambda\cdot\nabla)A_\mu^*\cdot\nabla)A_{\mu i}^* ]
 + {\rm c.c.} \biggr) \nonumber \\ 
&+& u_3 \biggl(A_{\lambda j}A_{\lambda j}((\nabla A_{\mu i}^*)\cdot(\nabla A_{\mu i}^*) +(\nabla\cdot A_\mu^*)(\nabla\cdot A_\mu^*)+\partial_k A_{\mu i}^*\partial_i A_{\mu k}^*)  + {\rm c.c.}\biggr) \biggr],
\label{eq:Sgrad4}
\end{eqnarray}

where 
\begin{eqnarray}
u_{1\pm}&=& {\overline I} \biggl[\frac{3}{10} \psi^{(5)}(y) {\rm ln}\biggl(1+\frac{{\overline I}}{3(1-{\overline I})} \biggr) + 2 \sum_{m > 0} \frac{1}{m^2} {\rm ln}\biggl(1+\frac{{\overline I}}{3(1-{\overline I} + \pi^2 T m/(4 E_{\rm F}) )} \biggr) \biggl(5 \psi^{(3)}(m+y) + \frac{84}{m^2} \psi^{(1)}(y+m) 
\nonumber \\
&+& \frac{24}{m}(\psi^{(2)}(y) - \psi^{(2)}(y+m)) + \frac{168}{m^3}(\psi(y) - \psi(y+m)) \biggr) \biggr] 
\pm \psi^{(4)}(y), \nonumber \\
u_{2\pm} &=&  {\overline I} \biggl[\frac{1}{10} \psi^{(5)}(y) {\rm ln}\biggl(1+\frac{{\overline I}}{3(1-{\overline I})} \biggr) + 2 \sum_{m > 0} \frac{1}{m^2} {\rm ln}\biggl(1+\frac{{\overline I}}{3(1-{\overline I} + \pi^2 T m/(4 E_{\rm F}) )} \biggr) \biggl(4 \psi^{(3)}(m+y) + \frac{168}{m^2} \psi^{(1)}(y+m) \nonumber \\
&+& \frac{36}{m}(\psi^{(2)}(y) - \psi^{(2)}(y+m)) + \frac{336}{m^3}(\psi(y) - \psi(y+m)) \biggr) \biggr] 
\pm \frac{1}{3} \psi^{(4)}(y), \nonumber \\
u_3 &=& \frac{1}{2} u_{2-} - \frac{1}{3} \psi^{(4)}(y). 
\label{eq:upmn}
\end{eqnarray}
\end{widetext}
Here, $\psi(y)$ is the di-gamma function, and $\psi^{(k)}(y)=d^k \psi(y)/d y^k$. The terms accompanied by the $m$-summation imply the contributions arising from the quantum SF with $\Omega=2 \pi m T \neq 0$ neglected in eq.(\ref{eq:delchidd}). The terms proportional to $\psi^{(4)}(y)$ are the contributions of the WC diagram Fig.1, and the remaining terms in eq,(\ref{eq:Sgrad4}) are the results from Fig.3 (c). By applying eq.(\ref{eq:opa2}) to eq.(\ref{eq:Sgrad4}) with keeping the amplitude $|\Delta|$ fixed, we find eq.(\ref{eq:difK2}) in the text. 

Finally, the detailed expression of the gradient energy of O($|\Delta|^4$) in the FL approach will be given. For simplicity, the Landau parameter $\Gamma_1^{a} \equiv F_1^a/(1+F_1^a/3)$ will be set to be zero, because $F_1^a$ is usually believed to be much smaller than $F_1^s$. Then, the sum of eq.(\ref{eq:Gorboxgrad}) and eq.(\ref{eq:FLgrad2}) becomes 

\begin{widetext}
\begin{eqnarray}
f_{FLgrad4} &=& \frac{N(0)\psi^{(4)}(y)}{26880 (\pi T)^2} \biggl(\frac{v_{\rm F}}{2 \pi T} \biggr)^2 \biggl[w_{1} \biggl( \, 
(\nabla\cdot A_\mu)(\nabla\cdot A_\lambda^*)A_{\mu i}^*A_{\lambda i} 
+ (\nabla A_{\mu i})\cdot(\nabla A_{\lambda j}^*)A_{\mu i}^*A_{\lambda j} 
+ (A_\lambda\cdot\nabla)A_{\lambda i}^*(A_\mu^*\cdot\nabla)A_{\mu i} \biggr) \nonumber \\ 
&+& A_{\lambda j}A_{\mu j}^*\partial_k A_{\lambda i}^*\partial_i A_{\mu k} 
+(\nabla\cdot A_{\mu})(\nabla\cdot A_{\mu}^*)A_{\lambda j}A_{\lambda j}^*  
+ A_{\lambda j}A_{\lambda j}^*\partial_k A_{\mu i}^*\partial_i A_{\mu k} 
+(\nabla A_{\mu i})\cdot(\nabla A_{\mu i}^*)A_{\lambda j}A_{\lambda j}^* \nonumber \\ 
&+& (\nabla A_{\mu i})\cdot(\nabla A_{\mu j}^*)(A_{\lambda i}A_{\lambda j}^*+A_{\lambda i}^*A_{\lambda j}) + (\nabla A_{\mu i})\cdot(\nabla A_{\lambda i}^*) A_{\mu j}^*A_{\lambda j} 
+ (\nabla A_{\mu i})\cdot(\nabla A_{\lambda j}^*)A_{\mu j}^*A_{\lambda i} 
\nonumber \\ 
&+& (A_{\lambda}\cdot\nabla)A_{\mu i}^* (A_{\lambda}^*\cdot\nabla)A_{\mu i} 
+ (A_\lambda\cdot\nabla)A_{\mu i}(A_{\lambda}^*\cdot\nabla)A_{\mu i}^* 
+ (A_\lambda\cdot\nabla)A_{\mu i}(A_\mu^*\cdot\nabla)A_{\lambda i}^* 
 \nonumber \\
&+& \biggl[ w_1 \biggl( (\nabla\cdot A_\mu)(A_{\mu i}^*(A_\lambda\cdot\nabla)A_{\lambda i}^* + A_{\lambda i}(A_{\mu}^*\cdot\nabla)A_{\lambda i}^*) + 
A_{\mu i}^*((A_\lambda\cdot\nabla)A_\lambda^*\cdot\nabla)A_{\mu i} \biggr) \nonumber \\
&+& (\nabla\cdot A_\mu)(A_{\lambda i}^*(A_\lambda\cdot\nabla)A_{\mu i}^* 
+ A_{\lambda i}(A_{\lambda}^*\cdot\nabla)A_{\mu i}^*) 
+ A_{\lambda i}((A_\mu^*\cdot\nabla)A_\lambda^*\cdot\nabla)A_{\mu i} 
+ A_{\lambda i}((A_{\lambda}^*\cdot\nabla)A_{\mu}^*\cdot\nabla)A_{\mu i} \nonumber \\ 
&+& A_{\lambda i}^*((A_{\lambda}\cdot\nabla)A_{\mu}^*\cdot\nabla)A_{\mu i} 
 + {\rm c.c.} \biggr] 
- \biggl((\nabla A_{\mu i})\cdot(\nabla A_{\lambda i}^*)A_{\mu j}A_{\lambda j}^* + (\nabla A_{\mu i})\cdot(\nabla A_{\lambda j}^*)(A_{\mu i}A_{\lambda j}^* 
+ A_{\mu j}A_{\lambda i}^*) \nonumber \\ 
&+& (\nabla\cdot A_\mu)(\nabla\cdot A_\lambda^*)A_{\mu i}A_{\lambda i}^* + A_{\mu i}A_{\lambda i}^*\partial_k A_{\lambda j}^*\partial_j A_{\mu k} 
+ (A_{\lambda}^*\cdot\nabla)A_{\mu i}(A_\mu\cdot\nabla)A_{\lambda i}^*  + (A_{\mu}\cdot\nabla)A_{\mu i}(A_{\lambda}^*\cdot\nabla)A_{\lambda i}^* \nonumber \\
&+& [ (\nabla\cdot A_\mu)(A_{\lambda j}^*(A_\mu\cdot\nabla)A_{\lambda j}^* 
+ A_{\mu j}(A_\lambda^*\cdot\nabla)A_{\lambda j}^*) 
+ A_{\mu i}((A_\lambda^*\cdot\nabla)A_\lambda^*\cdot\nabla)A_{\mu i} + A_{\lambda i}^*((A_\mu\cdot\nabla)A_\lambda^*\cdot\nabla)A_{\mu i} 
+ {\rm c.c.}] \biggr) \nonumber \\ 
&+& \frac{1}{3} \biggl( w_2 \biggl( (\nabla A_{\mu i}^*)\cdot(\nabla A_{\lambda j}^*)A_{\mu i}A_{\lambda j} + (A_{\mu}\cdot\nabla)A_{\mu i}^*(A_{\lambda}\cdot\nabla)A_{\lambda i}^* + (\nabla\cdot A_\mu^*)(\nabla\cdot A_\lambda^*)A_{\mu i}A_{\lambda i} \biggr) \nonumber \\ 
&+& (\nabla A_{\mu i}^*)\cdot(\nabla A_{\lambda j}^*)A_{\mu j}A_{\lambda i} + (\nabla A_{\mu i}^*)\cdot(\nabla A_{\lambda i}^*)A_{\mu j}A_{\lambda j} 
+ (A_{\lambda}\cdot\nabla)A_{\mu i}^*(A_{\mu}\cdot\nabla)A_{\lambda i}^* 
+ A_{\mu i}A_{\lambda i}\partial_j A_{\mu k}^*\partial_k A_{\lambda j}^* \nonumber \\ 
&+& 2 \biggl[ w_2 \biggl((\nabla\cdot A_\mu^*)(A_{\lambda i}(A_\mu\cdot\nabla)A_{\lambda i}^* 
+ A_{\mu i}(A_\lambda\cdot\nabla)A_{\lambda i}^*) +
A_{\lambda i}((A_\mu\cdot\nabla)A_\mu^*\cdot\nabla)A_{\lambda i}^* \biggr)+ A_{\lambda i}((A_\mu\cdot\nabla)A_\lambda^*\cdot\nabla)A_{\mu i}^*  \biggr]
 \nonumber \\ 
&+& {\rm c.c.} \biggr) 
- \frac{1}{3} \biggl( (\nabla A_{\mu i}^*)\cdot(\nabla A_{\mu j}^*)A_{\lambda i}A_{\lambda j} + (A_{\lambda}\cdot\nabla)A_{\mu i}^*(A_{\lambda}\cdot\nabla)A_{\mu i}^* 
+ 2[(\nabla\cdot A_\mu^*)A_{\lambda i}(A_\lambda\cdot\nabla)A_{\mu i}^* \nonumber \\ 
&+& A_{\lambda i}((A_\lambda\cdot\nabla)A_\mu^*\cdot\nabla)A_{\mu i}^* ]
 + {\rm c.c.} \biggr) - \frac{1}{2} \biggl(A_{\lambda j}A_{\lambda j}((\nabla A_{\mu i}^*)\cdot(\nabla A_{\mu i}^*) \nonumber \\
&+& (\nabla\cdot A_\mu^*)(\nabla\cdot A_\mu^*)+\partial_k A_{\mu i}^*\partial_i A_{\mu k}^*)  + {\rm c.c.}\biggr) \biggr],
\label{eq:FLgrad4}
\end{eqnarray}
\end{widetext}

where 
\begin{eqnarray}
w_1 &=& 1 +\frac{28}{15} \Gamma_1^{s} \frac{(\psi^{(2)}(y))^2}{\psi^{(4)}(y)}, \nonumber \\
w_2 &=& 1 - \frac{14}{5} \Gamma_1^{s} \frac{(\psi^{(2)}(y))^2}{\psi^{(4)}(y)}, 
\end{eqnarray}
and the terms proportional to $\psi^{(4)}(y)$ are the WC contributions from Fig.1.


\begin{thebibliography}{99}
\bibitem{com1} Here, a double (spatial) derivative of a single field has been rewritten by a partial integration into a form consisting of products of a single derivative of a field.
\bibitem{comexp} In contrast, entering the superfluid phase by a cooling followed by a subsequent rotation is expected to lead to avoiding emergence of PVs.
\end{thebibliography}
\end{document}